\documentclass[10pt,aps,prc,twocolumn,superscriptaddress,showpacs,floatfix,nofootinbib]{revtex4-1}

\usepackage{array}
\usepackage{amsmath}
\usepackage{mathptmx}
\usepackage{mathptm}
\usepackage{amssymb}
\usepackage{longtable}
\usepackage{bm}
\usepackage{wasysym}
\usepackage{multirow}
\usepackage{float}
\usepackage[final]{graphicx}
\usepackage{epstopdf}
\usepackage{placeins}
\usepackage{xcolor}
\usepackage{subfloat}
\usepackage{rotating}
\usepackage{booktabs}
\usepackage{tabularx}
\usepackage[caption=false]{subfig}
\usepackage[sort&compress]{natbib}
\usepackage{fontenc}
\usepackage{times}
\usepackage{epsfig}
\usepackage{verbatim}
\usepackage{threeparttable}
\usepackage{dcolumn}
\usepackage[english]{babel}
\usepackage{chngcntr}
\usepackage{hycolor}
\usepackage[colorlinks=true,linkcolor=blue,citecolor=blue,urlcolor=blue]{hyperref}
\usepackage{siunitx}
\usepackage{etoolbox}
\robustify\itshape

\makeatletter

\newcolumntype{.}{D{.}{.}{1}}
\newcolumntype{X}{D{X}{X}{1}}

\newcommand{\vast}{\bBigg@{2}} 
\newcommand{\vastp}{\bBigg@{3}} 
\newcommand{\Vast}{\bBigg@{3.5}}
\newcommand{\vastt}{\bBigg@{4}}
\newcommand{\Vastt}{\bBigg@{4.5}}
\makeatother

\AtBeginDocument{
\heavyrulewidth=.08em
\lightrulewidth=.05em
\cmidrulewidth=.03em
\belowrulesep=.65ex
\belowbottomsep=0pt
\aboverulesep=.4ex
\abovetopsep=0pt
\cmidrulesep=\doublerulesep
\cmidrulekern=.5em
\defaultaddspace=.5em
}

\begin{document}


\title{A study of $^{35}$Cl excited states via $^{32}$S($\alpha, p$)}
\author{K.~Setoodehnia}
 \altaffiliation{Present address: European X-Ray Free Electron Laser GmbH, Holzkoppel 4, 22869 Schenefeld, Germany}
 \email{kiana.setoodehnia@xfel.eu}
 \affiliation{Department of Physics, North Carolina State University, Raleigh NC, 27695, USA\\}
 \affiliation{Triangle Universities Nuclear Laboratory, Duke University, Durham NC, 27710, USA\\}
\author{J.~H.~Kelley}
 \affiliation{Department of Physics, North Carolina State University, Raleigh NC, 27695, USA\\}
 \affiliation{Triangle Universities Nuclear Laboratory, Duke University, Durham NC, 27710, USA\\}
\author{C.~Marshall}
 \affiliation{Department of Physics, North Carolina State University, Raleigh NC, 27695, USA\\}
 \affiliation{Triangle Universities Nuclear Laboratory, Duke University, Durham NC, 27710, USA\\}
\author{F.~Portillo Chaves}
 \affiliation{Department of Physics, North Carolina State University, Raleigh NC, 27695, USA\\}
 \affiliation{Triangle Universities Nuclear Laboratory, Duke University, Durham NC, 27710, USA\\}
\author{R.~Longland}
 \affiliation{Department of Physics, North Carolina State University, Raleigh NC, 27695, USA\\}
 \affiliation{Triangle Universities Nuclear Laboratory, Duke University, Durham NC, 27710, USA\\}


\date{\today}


\begin{abstract}
\textbf{Background:} The presolar grains originating in oxygen-neon novae may be identified more easily from those of other stellar sources if their sulfur isotopic ratios ($^{33}$S/$^{32}$S and $^{34}$S/$^{32}$S) are compared with the theoretical ones. The accuracy of such a comparison depends on reliable $^{33}$S($p, \gamma$)$^{34}$Cl and $^{34}$S($p, \gamma$)$^{35}$Cl reaction rates at the nova temperature regime. The latter rate has recently been computed based on experimental input, and many new excited states in $^{35}$Cl were discovered above the proton threshold. As a result, the experimental $^{34}$S($p, \gamma$)$^{35}$Cl rate was found to be less uncertain and 2 -- 5 times smaller than the theoretical one. Consequently, the simulated $^{34}$S/$^{32}$S isotopic ratio for nova presolar grains was predicted to be smaller than that of type II supernova grains by a factor of 1.5 to 3.7. \textbf{Purpose:} The present study was performed to confirm the existence of these new resonances, and to improve the remaining uncertainties in the $^{34}$S($p, \gamma$)$^{35}$Cl reaction rate. \textbf{Methods:} Energies and spin-parities of the $^{35}$Cl excited levels were investigated via high-resolution charged-particle spectroscopy with an Enge split-pole spectrograph using the $^{32}$S($\alpha, p$)$^{35}$Cl reaction. Differential cross sections of the outgoing protons were measured at $E_{\alpha}$ $=$ 21 MeV. Distorted-wave Born approximation calculations were carried out to constrain the spin-parity assignments of observed levels with special attention to those significant in determination of the $^{34}$S($p, \gamma$)$^{35}$Cl reaction rate over the nova temperature regime. \textbf{Results:} The existence of these newly discovered states are largely confirmed, although a few states were not observed in this study. The spins and parities of a few $^{35}$Cl states were assigned tentatively for the first time. \textbf{Conclusions:} The present $^{34}$S($p, \gamma$)$^{35}$Cl experimental thermonuclear reaction rate at 0.1 -- 0.4 GK is consistent within 1$\sigma$ with the previous evaluation. However, our rate uncertainty is larger than before due to a more realistic treatment of the uncertainties in the rate input. In comparison with the previous rate evaluation, where the high and low rates differed by less than a factor of 2 over nova temperature regime, the ratio of the present limit rates is at most a factor of 3.5 at 0.12 GK. At temperatures above 0.2 GK, we recommend the future work to focus on determination of the unknown properties of four excited states of $^{35}$Cl 6643 keV, 6761 keV, 6780 keV, and 6800 keV.
\end{abstract}


\maketitle


\section{\label{Introduction}Introduction}

Classical novae are the third most energetic stellar explosions in the universe. They are powered by a thermonuclear runaway, which is caused by the accretion of hydrogen-rich matter onto the surface of a white dwarf that is in a close binary system with a main sequence star. During a classical nova event and depending on the mass of the white dwarf, peak temperatures of 0.1 -- 0.4 GK are reached. At these elevated temperatures, nucleosythesis proceeds via the rp-process~\cite{Bertulani:2016}, and matter is synthesized up to $A$ $\sim$ 40 by explosive hydrogen burning through a series of ($p, \gamma$) and ($p, \alpha$) reactions and $\beta^{+}$-decays on the proton-rich side of the valley of stability.\par
Systematic infrared~\cite{Evans:1990,Gehrz:1998,Gehrz:1999,Amari:2001,Starrfield:2007} and ultraviolet~\cite{Shore:1994} observations of nova light curves reveal episodes of dust condensation and grain formation in the expanding shells of the nova ejecta. As the solar system was forming from a molecular cloud 4 billion years ago, these grains found their way into this cloud. These presolar grains carry non-solar isotopic signatures and are tiny samples of nucleosynthesis associated to the site where they were created. They are discovered through the laboratory analysis of primitive meteorites, which yields isotopic abundance ratios in these grains~\cite{Nittler:2016}. Such measurements in the presolar grains of nova origin can add powerful observational constraints on the theoretical nucleosynthesis predictions derived from the nuclear reaction networks used in nova evolution calculations.\par
A few candidate presolar grains of nova origin have been found~\cite{Amari:2001,Amari:2002}. They are characterized by rather large isotopic anomalies (compared to the solar values) that can be explained in terms of the imprints of nova nucleosynthesis (e.g., low $^{12}$C/$^{13}$C and $^{14}$N/$^{15}$N, high $^{30}$Si/$^{28}$Si and $^{22}$Ne/$^{20}$Ne ratios~\cite{Jose:2004,Jose:2007}). More recently, it has been suggested~\cite{Hoppe:2010,Gyngard:2012,Hoppe:2012,Parikh:2014} that measurements of $^{33}$S/$^{32}$S and $^{34}$S/$^{32}$S isotopic ratios, together with other nova isotopic signatures, in presolar grains can provide additional support in identifying presolar grains of oxygen-neon novae from those of type II supernovae~\cite{Gillespie:2017}. However, the $^{33}$S($p, \gamma$)$^{34}$Cl and $^{34}$S($p, \gamma$)$^{35}$Cl reaction rates must be known with sufficient accuracy over the nova temperature regime.\par
A sufficiently precise $^{33}$S($p, \gamma$)$^{34}$Cl reaction rate has been determined previously~\cite{Parikh:2014}. On the other hand, the $^{34}$S($p, \gamma$)$^{35}$Cl reaction rate was not known precisely enough due to uncertainties associated with estimation of a rate based on statistical models when the experimental information is scarce. This was the case until late--2017 when the results of the measurement of Ref.~\cite{Gillespie:2017} was published. This measurement is the first to reduce the uncertainty in the $^{34}$S($p, \gamma$)$^{35}$Cl reaction rate. The subsequently predicted $^{34}$S/$^{32}$S isotopic ratio from an oxygen-neon nova simulation~\cite{Gillespie:2017} was estimated to be about a factor of 2 to 3 lower than that from recent models of a type II supernova.\par
The $^{34}$S($p, \gamma$)$^{35}$Cl reaction ($Q$-value $=$ 6370.81(4) keV~\cite{Wang:2017}) rate over the temperature range corresponding to explosive hydrogen burning in novae is dominated by contributions from the $^{35}$Cl excited states with 6493 keV $\lessapprox$ $E_{x}$ $\lessapprox$ 6927 keV.\par
Prior to the measurement of Ref.~\cite{Gillespie:2017}, the excited states of $^{35}$Cl had been previously measured using a variety of indirect methods such as transfer reactions, as well as a few direct measurements of $^{34}$S($p, \gamma$)$^{35}$Cl~\cite{Chen:2011} (and references therein). However, the energy of excited states in the range of interest remained poorly constrained, and the spin-parities of these states were either unknown or tentatively known. The high resolution measurement of Gillespie \textit{et al.}~\cite{Gillespie:2017} not only improved the $^{35}$Cl excitation energy uncertainties but ten previously unobserved states were also discovered. However, the spin-parities of the levels of interest still remained mostly tentative.\par
We performed an independent high-resolution charged-particle spectroscopy experiment via the $^{32}$S($\alpha, p$)$^{35}$Cl reaction. We specifically explored the $E_{x}$($^{35}$Cl) $\sim$ 6 -- 7 MeV region to confirm the energies and spin-parities of the astrophysically significant proton resonances in $^{35}$Cl.

\section{\label{Experiment}Experimental Setup}

The 10-MV FN tandem Van de Graaff accelerator at \underline{T}riangle \underline{U}niversities \underline{N}uclear \underline{L}aboratory (TUNL) accelerated a $^{4}$He$^{2+}$ beam to 21 MeV ($\Delta\,E$/$E$ $\sim$ 3.5 $\times$ 10$^{-4}$). Two high resolution 90$^{\circ}$ dipole magnets were used to analyze the beam energy and deliver the 1 mm (in diameter) beam to target. Typical beam intensity on target varied between 40 to 500 enA.\par
The $^{32}$S($\alpha, p$)$^{35}$Cl reaction was measured using antimony sulfide and cadmium sulfide targets. A silicon dioxide and a carbon target were also employed for calibration purposes and background determination, respectively. Except the carbon foil which was bought from the Arizona Carbon Foil Company~\cite{ACF-Metals}, the other targets were fabricated prior to the experiment by thermal vacuum evaporation of Sb$_{2}$S$_{3}$, CdS, and SiO$_{2}$ powders onto carbon foil substrates with various thicknesses. The thickness of the evaporated layers were monitored during the evaporation using a quartz crystal thickness monitor.\par
Except for the antimony sulfide target, the thickness and stoichiometry of each of the remaining targets were independently determined via a \underline{R}utherford \underline{b}ackscattering \underline{s}pectrometry (RBS) measurement following the main ($\alpha, p$) experiment. For the former, a 2-MeV $^{4}$He$^{2+}$ beam was employed using the same accelerator facility. A single 100-$\mu$m-thick silicon surface barrier detector was placed at 165$^{\circ}$ with respect to the beam axis to measure the backscattered $\alpha$-particles with 17-keV energy resolution. A pulser was used to adjust the gain of the silicon detector and monitor the dead time during the RBS measurement. The RBS spectra were energy calibrated using a gold target with a known thickness. The analysis of the RBS data resulted in the following contents for each target:\\
(i) The CdS target: 15.9 $\mu$g/cm$^{2}$ of $^{nat}$S, 43.6 $\mu$g/cm$^{2}$ of $^{nat}$Cd, and 31.9 $\mu$g/cm$^{2}$ of $^{nat}$C. (ii) The SiO$_{2}$ target: 14.7 $\mu$g/cm$^{2}$ of $^{nat}$Si, 30.2 $\mu$g/cm$^{2}$ of $^{nat}$O, 12 $\mu$g/cm$^{2}$ of $^{nat}$C, and 6.6 $\mu$g/cm$^{2}$ of $^{nat}$Ta, where the latter contamination comes from partial melting of the Ta evaporation boat towards the end of the evaporation. But no excited states from the tantalum contamination in the targets were observed. (iii) The C target: 30.2 $\mu$g/cm$^{2}$ of $^{nat}$C.\par
The antimony sulfide target degraded substantially (and suddenly) during the main ($\alpha, p$) experiment after about 54 hours of beam on target (21-MeV $^{4}$He at $\sim$ 250 enA). Its thickness was not confirmed by an independent RBS measurement. A spare Sb$_{2}$S$_{3}$ target evaporated at the same time had 46.3 $\mu$g/cm$^{2}$ of $^{nat}$S, 117.4 $\mu$g/cm$^{2}$ of $^{nat}$Sb, and 22.9 $\mu$g/cm$^{2}$ of $^{nat}$C. In the beginning of the experiment, the antimony sulfide target was utilized. After its degradation, the experiment was continued using the thinner CdS target instead since CdS is less susceptible to degradation. For consistency check, the $^{32}$S($\alpha, p$) reaction was measured with both targets at 30$^{\circ}$, and the resulting differential cross sections for different excited states of $^{35}$Cl at that angle were in agreement.\par
The uncertainties in the thicknesses of CdS, spare Sb$_{2}$S$_{3}$ and SiO$_{2}$ targets measured by RBS were taken to be $\approx$ 10\%, which is a conservative estimate of the uncertainty of stopping powers of helium in these materials, where no experimental data are available~\cite{SRIM}. The uncertainty in the thicknesses of the carbon target was determined to be 5.6\% from Ref.~\cite{SRIM2}. The CdS and both Sb$_{2}$S$_{3}$ targets were fairly clean and did not show any oxygen contamination.\par
The light reaction products from the interaction of the beam with the targets were separated according to their momenta by the TUNL high resolution Enge split-pole magnetic spectrograph~\cite{Setoodehnia:2016}. The magnetic field and the solid angle acceptance of the spectrograph were set to 0.67 T and 1 msr, respectively. The reaction products were measured at laboratory angles of 10${^\circ}$, 15${^\circ}$, 19${^\circ}$, 30${^\circ}$, 35${^\circ}$, 40${^\circ}$, 45${^\circ}$ and 50${^\circ}$. For $\theta_{\mbox{lab}}$ $=$ 10$^{\circ}$, $\Delta\Omega$ $=$ 0.5 msr was chosen to reduce (i) the background arising from scattered beam, and (ii) the detector deadtime from otherwise high count rates. Scattering angles between 20${^\circ}$ and 30${^\circ}$ were not considered because of an unexpected vacuum leak in the sliding seal which separates the target chamber and the spectrograph. Also, the region of interest would have been mostly obscured by the contaminant $^{1}$H($\alpha, p)$ reaction at these angles.\par
The spectrograph also focused the light reaction products onto its focal plane, where a high resolution position sensitive focal plane detector~\cite{Marshall:2018} detected particles whose radii of curvature were between 68 to 84 cm. This detector measured energy losses, residual energies and positions of the light reaction products along the focal plane of the spectrograph.\par
\begin{figure*}[ht]
 \begin{center}
  \subfloat{%
   \includegraphics[width=\textwidth]{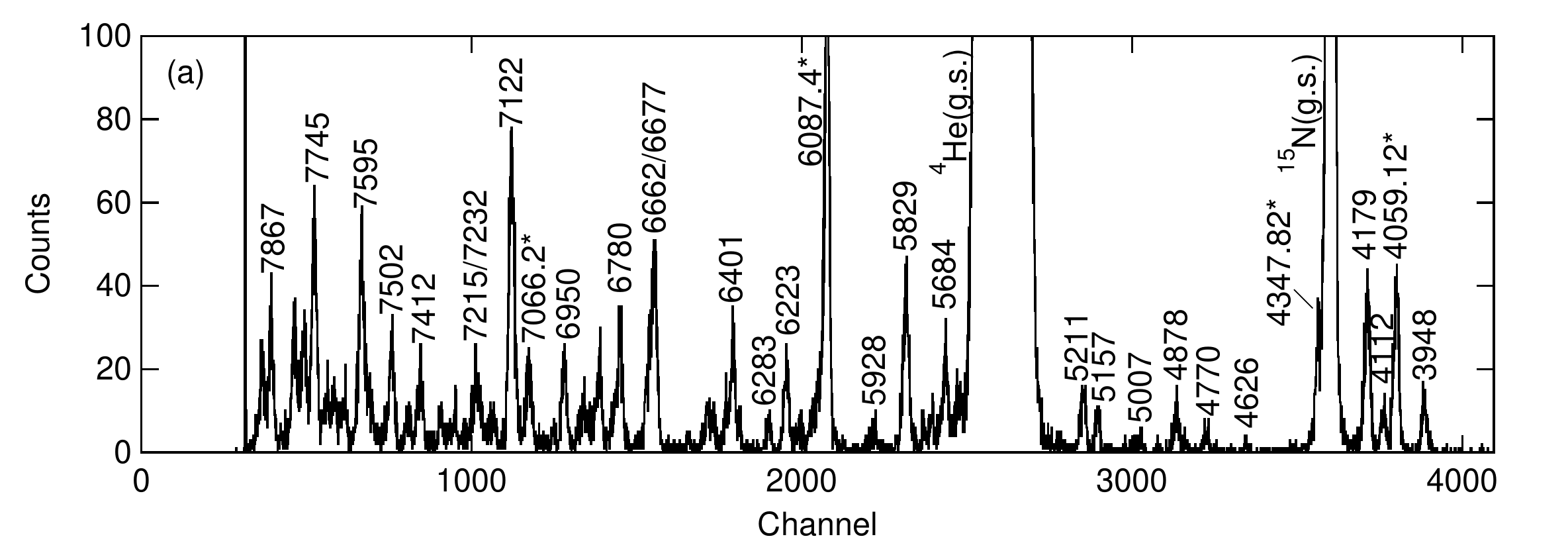}
   }\\
   \subfloat{%
   \vspace{-0.8mm}\includegraphics[width=\textwidth]{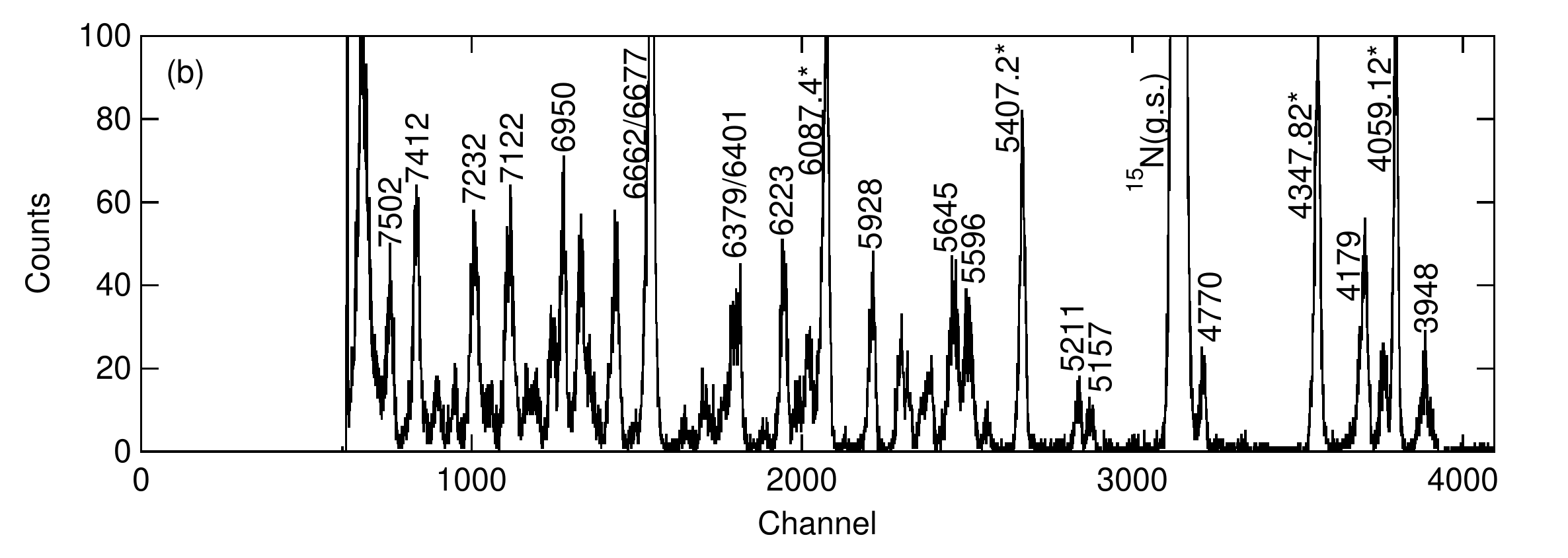}
   }\\
   \subfloat{%
   \vspace{-0.8mm}\includegraphics[width=\textwidth]{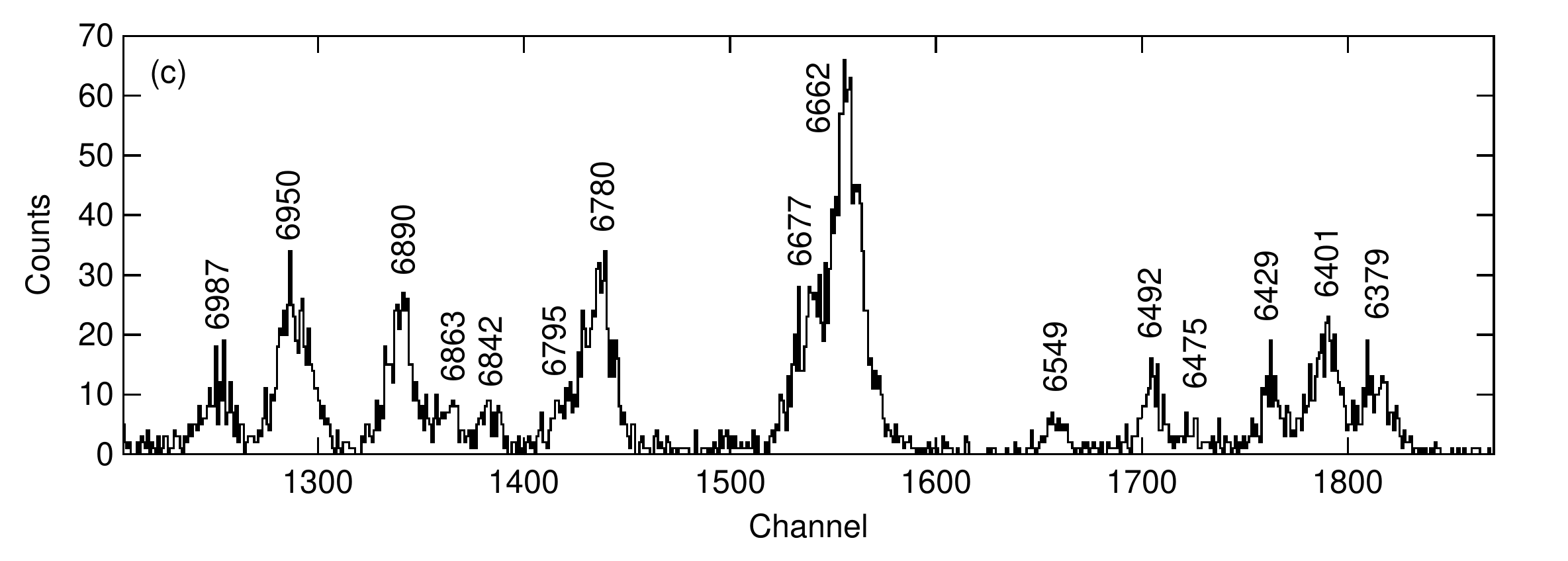}
   }
  \end{center}\vspace{-0.5cm}
\caption{\label{figure1}Spectra from the $^{32}$S($\alpha, p$)$^{35}$Cl reaction at $\theta_{lab}$ $=$ 15$^{\circ}$ (a), 40$^{\circ}$ (b) and 30$^{\circ}$ (zoomed in on the region of interest) (c). The spectra at 15$^{\circ}$ and 30$^{\circ}$ are obtained using the CdS target, and the one at 40$^{\circ}$ is obtained using the Sb$_{2}$S$_{3}$ target. The 40$^{\circ}$ spectrum is shifted back to compensate for the kinematics shift due to a change in the scattering angle. Therefore, the peaks in panels (a) and (b) are lined up with each other. Peaks corresponding to $^{35}$Cl states are labeled with energies (in keV) from the present work except those denoted by asterisks, which were used as internal calibration using energies from Ref.~\cite{Chen:2011}. For clarity, not all peaks are labeled. The main contaminant peaks are from the ground states (g.~s.) of $^{15}$N and $^{4}$He from the $^{12}$C($\alpha, p$) and $^{1}$H($\alpha, p$) reactions, respectively. The $^{4}$He(g.~s.) is significantly out of focus and broad due to the substantial differences in the kinematics of the $^{1}$H($\alpha, p$) and $^{32}$S($\alpha, p$) reactions.}
\end{figure*}

At each spectrograph angle, this information was used to obtain the momentum spectrum for protons from the $^{32}$S($\alpha, p$)$^{35}$Cl reaction corresponding to excited states in $^{35}$Cl (see Fig.~\ref{figure1}). These spectra were mostly free of contaminants except the ground state of $^{15}$N from the $^{12}$C($\alpha, p$) reaction occurring on the carbon substrates of the CdS and Sb$_{2}$S$_{3}$ targets, as well as the ground state of $^{4}$He from the $^{1}$H($\alpha, p$) reaction. The latter contamination had also been observed in the previ-

\begin{figure*}[ht]
\begin{center}
\includegraphics[width=\textwidth]{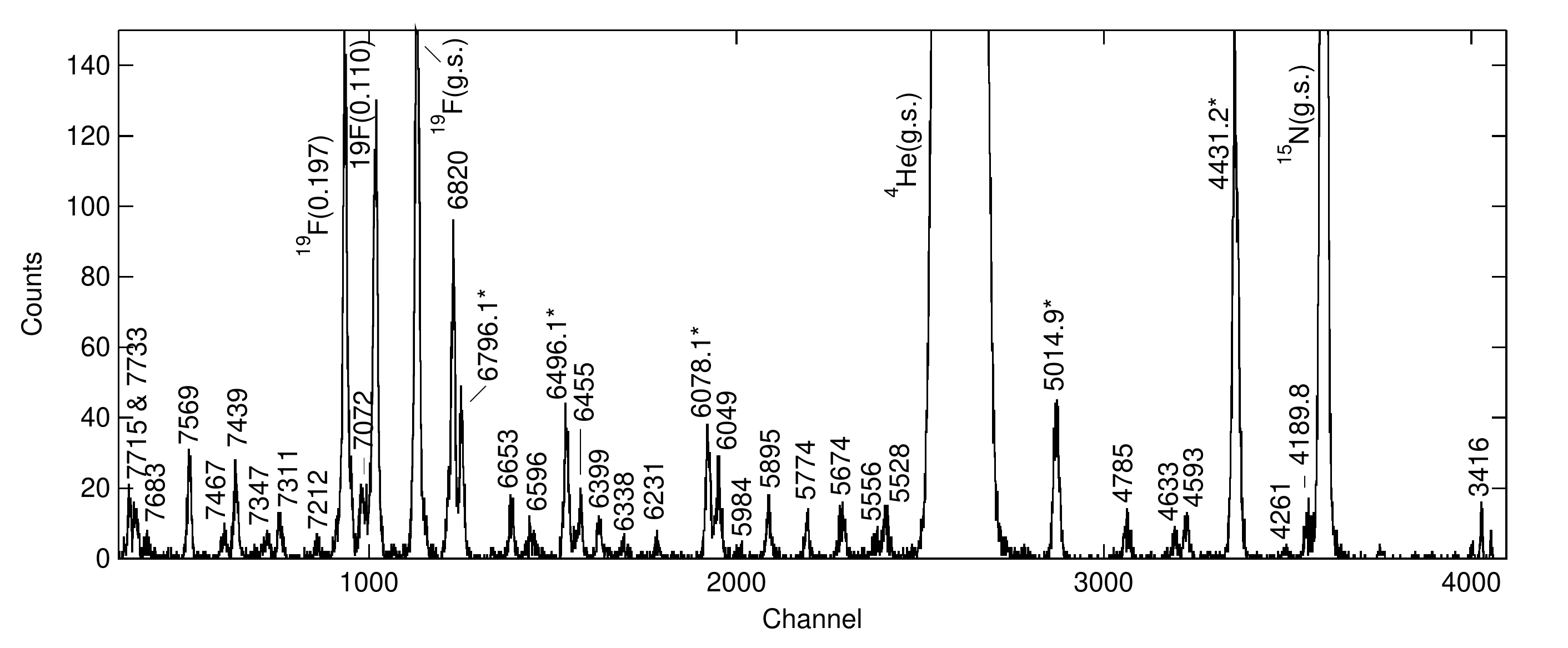}\\
\end{center}\vspace{-0.5cm}
\caption{\label{figure2}Spectrum from the $^{28}$Si($\alpha, p$)$^{31}$P calibration reaction measured using the SiO$_{2}$ target at $\theta_{lab}$ $=$ 15$^{\circ}$. Peaks corresponding to $^{31}$P states are labeled with energies (in keV, rounded to the nearest integer) from the present work. Those labeled by asterisk are used as calibration energies in our initial calibration fits. For the latter, the energies are adopted from Ref.~\cite{Ouellet:2013}. The main contaminant peaks are from the $^{1}$H($\alpha, p$)$^{4}$He, $^{12}$C($\alpha, p$)$^{15}$N, and $^{16}$O($\alpha, p$)$^{19}$F reactions and are labeled with their parent nuclei and their energies (in MeV). g.~s.~indicates ground state.}
\end{figure*}

\noindent ious $^{32}$S($\alpha, p$) measurement~\cite{Goss:1973}.

\section{\label{Analysis}Data Analysis}

A least-squares multi-Gaussian fit function was used to derive the centroids, widths, and areas of the observed spectral peaks corresponding to the $^{35}$Cl and $^{31}$P excited states produced from the $^{32}$S($\alpha, p$)$^{35}$Cl and $^{28}$Si($\alpha, p$)$^{31}$P reactions, respectively.
The Bayesian framework described in Ref.~\cite{Marshall:2018} was used together with the known levels of $^{31}$P~\cite{Ouellet:2013} measured using the SiO$_{2}$ target (see Fig.~\ref{figure2}) to initially identify and calibrate the well populated states on the $^{35}$Cl spectra. Once a good initial calibration fit was obtained, each $^{35}$Cl spectrum was recalibrated internally using the well populated well known states of $^{35}$Cl, whose energies were adopted from Ref.~\cite{Chen:2011} and are marked by asterisks in Table~\ref{tab:1}. All of the final internal calibrations were quadratic polynomial fits.\par
The uncertainties in excitation energies reported in Table~\ref{tab:1} arise from a convolution of the statistical uncertainties in the corresponding peak centroids; uncertainties in the coefficients of the polynomial calibration fits; and the reproducibility of the energies of the calibration peaks.\par
The systematic uncertainties in the $^{35}$Cl excitation energies obtained at each angle are mutually independent of those described above and were computed from: $\pm$10\% uncertainties in the thicknesses of the CdS, Sb$_{2}$S$_{3}$, and SiO$_{2}$ targets affecting energy losses through these targets; and the systematic uncertainty in the $Q$-value of the $^{32}$S($\alpha, p$) reaction, which is 0.04 keV~\cite{Wang:2017}. The uncertainty in the $Q$-value of the $^{28}$Si($\alpha, p$) reaction is negligible~\cite{Wang:2017}. A quadratic sum of these uncertainties results in an overall systematic uncertainty of 2 keV in each $^{35}$Cl excitation energy. This should be added in quadrature to the uncertainties quoted in Table~\ref{tab:1}. The final $^{35}$Cl excitation energies from the present work (listed in Table~\ref{tab:1}) are weighted average energies for each state over all the angles. These weighted average energies were computed using the V.AveLib utility code of Ref.~\cite{Birch:2012}.\par
The energy resolution defined as the peak \underline{f}ull \underline{w}idth at \underline{h}alf \underline{m}aximum (FWHM) was 24 keV averaged over all angles.

\section{\label{Results}Results}

The $^{35}$Cl states observed in the present work are listed in Table~\ref{tab:1}. All of these states have been observed at a minimum of three angles. Most of the measured energies in the present work are in agreement within 1 -- 2$\sigma$ with those measured in the previous $^{32}$S($\alpha, p$)$^{35}$Cl experiment~\cite{Goss:1973}, and the most recent measurement of the $^{34}$S($^{3}$He, $d$)$^{35}$Cl reaction~\cite{Gillespie:2017}, as well as with the excitation energies reported in the most recent evaluation of $^{35}$Cl excited states~\cite{Chen:2011} (and references therein). A few exceptions are the present 5531-, 5731-, 6475- and 6662-keV states. These are mostly states that are populated in a region with a high density of states, where the peaks are not too strongly populated (see Fig.~\ref{figure1}). Therefore, multiple peaks were fitted at once to obtain the peak properties. In order to achieve the best fits, the widths of these states sometimes had to be kept fixed to the average width of $^{35}$Cl states at that angle.\par
Furthermore, it is worth mentioning here that we have not observed a new state at $E_{x}$ $=$ 6643 keV, which was first measured in Ref.~\cite{Gillespie:2017}. Resolving this discrepancy (see \S~\ref{Rate}) proves to be significant in the determination of the $^{34}$S($p, \gamma$)$^{35}$Cl reaction rate over the nova temperature regime.\par

\LTcapwidth=\textwidth
\begin{longtable*}{lccccccclc}
\caption{\label{tab:1}Weighted average (over all angles) excitation energies (in keV) of $^{35}$Cl from the present work in comparison with the most recent evaluation of $^{35}$Cl excited states~\cite{Chen:2011} and the results of the previous $^{32}$S($\alpha, p$)~\cite{Goss:1973} and $^{34}$S($^{3}$He, $d$)~\cite{Gillespie:2017} measurements. States used in the present work for internal energy calibration are denoted by an asterisk and their energies are adopted from Ref.~\cite{Chen:2011}. The uncertainties reported here for the present work do not include the $\pm$2 keV systematic uncertainty in our results.}\\
\toprule[1.0pt]\addlinespace[0.6mm]
\multicolumn{2}{c}{$^{35}$Cl Evaluation~\cite{Chen:2011}} & \multicolumn{1}{c}{\phantom{ab}} & \multicolumn{2}{c}{$^{34}$S($^{3}$He, $d$)~\cite{Gillespie:2017}} & \multicolumn{1}{c}{\phantom{ab}} & \multicolumn{1}{c}{$^{32}$S($\alpha, p$)~\cite{Goss:1973}} & \multicolumn{1}{c}{\phantom{ab}} & \multicolumn{2}{c}{Present Work}                                                                                                                         \\
\cmidrule[0.05em]{1-2} \cmidrule[0.05em]{4-5}\cmidrule[0.05em]{7-7} \cmidrule[0.05em]{9-10}\addlinespace[0.3mm]
\multicolumn{1}{l}{$E_{x}$ (keV)} & \multicolumn{1}{c}{$J^{\pi}$} & \multicolumn{1}{c}{} & \multicolumn{1}{c}{$E_{x}$ (keV)} & \multicolumn{1}{c}{$\ell$} & \multicolumn{1}{c}{} & \multicolumn{1}{c}{$E_{x}$ (keV)} & \multicolumn{1}{c}{} & \multicolumn{1}{l}{$E_{x}$ (keV)} & \multicolumn{1}{c}{$J^{\pi}$}\\
\midrule \hline\addlinespace[0.6mm]
\endfirsthead
\multicolumn{10}{c}%
{{\bfseries \tablename\ \thetable{} -- continued from previous page}}                                                                                                                                                                                   \\ \toprule[1.0pt]\addlinespace[0.6mm]
\multicolumn{2}{c}{$^{35}$Cl Evaluation~\cite{Chen:2011}} & \multicolumn{1}{c}{\phantom{ab}} & \multicolumn{2}{c}{$^{34}$S($^{3}$He, $d$)~\cite{Gillespie:2017}} & \multicolumn{1}{c}{\phantom{ab}} & \multicolumn{1}{c}{$^{32}$S($\alpha, p$)~\cite{Goss:1973}} & \multicolumn{1}{c}{\phantom{ab}} & \multicolumn{2}{c}{Present Work}                                                                                                                         \\
\cmidrule[0.05em]{1-2} \cmidrule[0.05em]{4-5}\cmidrule[0.05em]{7-7} \cmidrule[0.05em]{9-10}\addlinespace[0.3mm]
\multicolumn{1}{l}{$E_{x}$ (keV)} & \multicolumn{1}{c}{$J^{\pi}$} & \multicolumn{1}{c}{} & \multicolumn{1}{c}{$E_{x}$ (keV)} & \multicolumn{1}{c}{$\ell$} & \multicolumn{1}{c}{} & \multicolumn{1}{c}{$E_{x}$ (keV)} & \multicolumn{1}{c}{} & \multicolumn{1}{l}{$E_{x}$ (keV)} & \multicolumn{1}{c}{$J^{\pi}$}\\
\midrule\hline\addlinespace[0.6mm]
\endhead
\hline \multicolumn{10}{c}{{Continued on next page}}                                                                                                                                                                                                     \\ \hline
\endfoot
\endlastfoot
3943.82(25)	     & 9/2$^{+}$ 		    &&	       &	     && 3943.7(23) && 3947.9(24) &  \\
4059.12(15)	     & 3/2$^{-}$ 		    &&	       &	 	 && 4056.9(27) && 4059.12*   & 3/2$^{-}$ \\
4111.98(24)	     & 7/2$^{+}$ 		    &&	       &	 	 && 4110.2(29) && 4112.0(23) & 7/2$^{+}$ \\
4177.88(15)	     & 3/2$^{-}$ 	 	    &&	       &	 	 && 4177.5(24) && 4179(3)    & 3/2$^{-}$ \\
4347.82(15)	     & 9/2$^{-}$ 	 	    &&	       &	 	 && 4346.6(24) && 4347.82*   &  \\
4624.35(23)	     & (3/2,5/2$^{+}$)      &&	       &	 	 && 4624.4(31) && 4626.1(20) &  \\
4768.82(18)	     & 7/2			        &&	       &	 	 && 4770.8(26) && 4769.9(20) &  \\
4854.4(4)	     & (1/2,3/2)	 	    &&	       &	 	 && 	       && 4859.2(19) &  \\
4881.07(21)	     & 7/2		 	        &&	       &	 	 && 4883.1(29) && 4878(4)    &  \\
5010.09(20)	     & (1/2,3/2)	 	    &&	       &	 	 &&	    	   && 5006.9(18) &  \\
5157(11)	     & 3/2$^{+}$,5/2$^{+}$ 	&&	       &	 	 && 5161.7(33) && 5156.8(21) & 5/2$^{+}$ \\
5215.79(18)	     & (3/2$^{+}$,5/2)		&&	       &	 	 && 5206.6(37) && 5211.1(20) & (5/2$^{+}$) \\
5403.5(3)	     & 1/2$^{-}$,3/2$^{-}$ 	&&	       &	 	 && 5402.0(29) &&            &  \\
5407.2(4)	     & 11/2$^{-}$ 		    &&	       &	 	 &&            && 5407.2*    & 11/2$^{-}$ \\
5520.0(11)	     &  			        &&	       &	 	 && 	       && 5531(4)    &  \\
5586.0(3)	     & 5/2$^{+}$		    &&	       &	 	 && (5576)	   && 5586.0(16) &  \\
5599.69(23)	     & 3/2$^{+}$,5/2$^{+}$	&&	       &	 	 && 5591.8(32) && 5596.5(17) &  \\
5633(3)	      	 &  			        &&	       &	 	 && 5633.1(32) && 5634.1(19) &  \\
5645.0(3)	     & (5/2,7/2,9/2$^{+}$) 	&&	       &	 	 && 	       && 5645(3)    &  \\
5654.48(22)	     & 3/2$^{+}$ 		    &&	       &	 	 && 	       && 5653(3)    &  \\
5682.9(6)	     & 1/2$^{-}$,3/2$^{-}$ 	&&	       &	 	 && 5677.7(34) && 5684(4)    &  \\
5723.6(4)	     & 5/2$^{+}$ 		    &&	       &	 	 &&	    	   && 5731.2(15) &  \\
5758.0(4)	     & (1/2$^{+}$,3/2) 		&&	       &	 	 &&	    	   && 5757(3)    &  \\
5805.5(4)	     & (1/2$^{+}$,3/2,5/2) 	&&	       &	 	 && 5809.2(34) && 5807.1(24) &  \\
5823.0(10)	     & (5/2,9/2) 		    &&	       &	 	 && (5823)	   && 5829(3)    & 5/2$^{(-)}$ \\
5926.9(3)	     & 11/2$^{-}$ 		    &&	       &	 	 && 5927.4(35) && 5928.2(19) &  \\
6087.4(4)	     & 13/2$^{-}$ 		    &&	       &		 && 6084.2(29) && 6087.4*    & 13/2$^{-}$ \\
6106.2(4)	     & (3/2,5/2$^{+}$) 		&&	       &	 	 &&	    	   && 6104(3)    &  \\
6139(4)	      	 & 5/2$^{+}$ 		    &&	       &	 	 && 6140.2(40) && 6142(3)    &  \\
6181.0(6)	     & (1/2:7/2,9/2$^{-}$) 	&&	       &	 	 &&	    	   && 6180.4(22) &  \\
6225(4)	         &  			        &&	       &	 	 && 6224.9(36) && 6223.5(25) &  \\
		         &  			        && 6284(4) & 2	     &&	    	   && 6282.6(17) & (5/2$^{+}$) \\
		         &  			        && 6329(4) & 0/1     &&	    	   &&	         &  \\
6380.8(8)	     &  			        && 6377(2) & 2/3     && 6379.0(34) && 6379.3(14) & (9/2$^{-}$) \\
6402(4)	         &  			        &&	       & 	     && 6402.4(41) && 6400.9(10) & (1/2$^{-}$) \\
		         &  			        && 6427(2) & 3	 	 && (6427)	   && 6428.6(19) & (1/2$^{+}$) \\
		         &  			        && 6468(2) & 1	 	 &&	    	   && 6475(3)\footnote{\label{footnote1}No DWBA calculation was performed due to lack of enough angular data.}    &  \\
6492.0(6)	     & (1/2,3/2,5/2$^{+}$) 	&& 6491(2) & 2	 	 && 6491.9(34) && 6491.8(21) & (3/2$^{+}$) \\
		         &     			        && 6545(2) & 0/1	 &&	    	   && 6548.8(24) & (1/2$^{+}$) \\
      	         &  			        && 6643(2) & 1	 	 &&            &&            &  \\
6656(3)	         &  			        &&         &         && 6656.0(31) && 6662.2(19) & (7/2$^{+}$) \\
6681(3)	         &  			        && 6674(2) & 1/2/3	 && 6680.8(31) && 6677(3)    & 1/2$^{+}$ \\
6746(12)	     & 3/2$^{+}$,5/2$^{+}$ 	&& 6761(2) & 0/1	 &&	    	   &&	           &  \\
6783(3)	         &  			        && 6778(2) & 1	 	 && 6782.8(32) && 6779.8(20) & (3/2$^{-}$) \\
6802(4)	         &  			        &&	       &	 	 && 6802.1(42) && 6795(6)$^{\ref{footnote1}}$    &  \\
		         &  			        && 6823(2) & 1	 	 &&	    	   &&            &  \\
		         &  			        && 6842(2) & 2/3	 &&	    	   && 6842(3)    & (3/2$^{+}$) \\
6866.7(6)	     &  			        && 6866(2) & 0 $+$ 2 && (6867)	   && 6863.1(21) & (9/2$^{+}$) \\
6894(3)	         &  			        &&	       &	 	 && 6893.5(32) && 6890.4(22) & (9/2$^{+}$) \\
6947(4)	         & 5/2$^{+}$ 		    &&	       &	 	 && 6947.5(34) && 6950(3)    & 5/2$^{+}$ \\
6986(4)	         &  			        &&	       &	 	 &&            && 6987(3)    & (9/2$^{+}$) \\
7066.2(3)	     & 5/2$^{+}$ 		    && 7066(2) & 1/2	 &&	    	   && 7066.2*    & 5/2$^{+}$ \\
7103.3(3)	     & 3/2 			        && 7103(2) & 1/3	 &&      	   &&            &  \\
7121(4)	         &  			        &&	       &	 	 &&            && 7122.1(13) & (5/2$^{-}$) \\
7170(10)         & (7/2 : 17/2)$^{+}$   && 7178(2) & 2       &&            && 7180(3)    & (7/2$^{+}$) \\
7185.0(3)        & 5/2$^{+}$            &&         &         &&            &&            &  \\
7194.5(3)	     & 1/2$^{-}$ 		    && 7194(2) &         &&	    	   &&	         &  \\
7210(4)	         &  			        &&	       &	 	 &&            && 7215.2(14)$^{\ref{footnote1}}$ &  \\
7225.5(3)        & 5/2       			&& 7227(2) & 0/1	 &&            &&            &  \\
7234.0(3)        & 5/2$^{+}$ 			&& 7227(2) & 0/1	 &&            && 7231.6(20) & (3/2$^{+}$) \\
7269.2(1)/7272.6(3)	     & - / 1/2$^{-}$ 		    && 7273(2) & 0/1	 &&	    	   &&            &  \\
7348(5)	         &  			        &&	       &	 	 &&            && 7347.9(18) & (7/2)\footnote{The reduced $\chi^{2}$ for the best DWBA fits are as follows: $J^{\pi}$ $=$ 7/2$^{+}$: $\chi^{2}$/$\nu$ $=$ 3.55, $J^{\pi}$ $=$ 7/2$^{-}$: $\chi^{2}$/$\nu$ $=$ 3.53 (see \S~\ref{Spin-Parities}).} \\
7362.0(3)	     & 3/2 			        && 7361(2) & 1	 	 &&	    	   && 7362.4(23)$^{\ref{footnote1}}$ &  \\
7396.0(3)	     & 7/2$^{(-)}$ 		    && 7398(2) & 2/3	 &&	    	   &&	         &  \\
7418(5)	         &  			        &&	       &	 	 &&            && 7411.6(24) &  \\
7451.0(5)	     & 3/2 			        &&	       &	 	 &&	    	   && 7446.6(19) &  \\
7501.1(5)/7502.9(3)&  			        &&	       &	 	 &&            && 7502(5)    &  \\
7561.1(4)	     & (1/2,3/2) 		    &&	       &	 	 &&            && 7564(4)    &  \\
7587(4)/7600.8(3)& - / 5/2$^{+}$ 		&&	       &	 	 &&            && 7595(6)    &  \\
7650(4)	         &  			        &&	       &	 	 &&            && 7647.1(19) &  \\
7670(10)/7671.9(3)& (7/2:17/2)$^{+}$/(5/2$^{-}$,7/2)&&  &	 && 	       && 7674.6(15) &  \\
7706.4(3)	     & 5/2$^{+}$ 		    &&	       &	 	 &&	    	   && 7709.9(15) &  \\
7744.8(4)        & 7/2$^{-}$            &&         &		 &&            && 7744.6(25) &  \\
7750(10)	     & (7/2:17/2)$^{+}$     &&	       &	 	 &&            && 7768.5(14) &  \\
7796.6(4)	     & 1/2$^{-}$ 		    &&	       &	 	 &&            && 7798.7(26) &  \\
7868.6(5)/7873.2(4)& (3/2,5/2$^{+}$)/13/2$^{+}$&&  &	 	 &&            && 7867(7)    &  \\
7889.0(15)/7899.1(3)& - /(3/2$^{-}$,5/2)&&	       &	 	 &&	    	   && 7899(12)   &  \\[0.7ex]
\bottomrule[1.5pt]
\end{longtable*}

In the measurement of Ref.~\cite{Gillespie:2017}, 10 new $^{35}$Cl states had been observed. Except for the 6329-keV, 6643-keV, and 6823-keV states~\cite{Gillespie:2017}, which have remained unobserved in the present work, we have confirmed the existence of all the other newly discovered $^{35}$Cl levels.\par
Spin parity assignments for the states observed here are made through a comparison between the measured angular distributions of the center-of-mass differential cross sections of protons from the $^{32}$S($\alpha, p$) reaction and their theoretical counterparts computed via \underline{d}istorted \underline{w}ave \underline{B}orn \underline{a}pproximation (DWBA) calculations.

\subsection{\label{Scattering}$^{4}$He $+$ $^{32}$S Elastic scattering measurement}

To obtain the entrance channel optical model parameters used in the DWBA calculations, we measured $^{4}$He $+$ $^{32}$S elastic scattering at $E_{\alpha}$ $=$ 21-MeV. This was measured at $\theta_{lab}$ $=$ 20$^{\circ}$, 22$^{\circ}$, 30$^{\circ}$, 35$^{\circ}$, 40$^{\circ}$, 45$^{\circ}$, and 50$^{\circ}$ using the CdS target, and the TUNL Enge split-pole spectrograph together with its focal plane detector package. At $\theta_{lab}$ $<$ 20$^{\circ}$, carbon and sulfur elastic scattering peaks were unresolved, and therefore, were not considered.\par
For this measurement, beam intensity was $\sim$ 100 enA. The magnetic field of the spectrograph was set to 0.85 T. The solid angle acceptance of the spectrograph was chosen to be 0.5 msr for all angles except 45$^{\circ}$ and 50$^{\circ}$, where $\Delta\Omega$ was changed to 2 msr to increase the count rates of the scattering events.\par
\begin{figure*}[ht]
\begin{center}
\includegraphics[width=\textwidth]{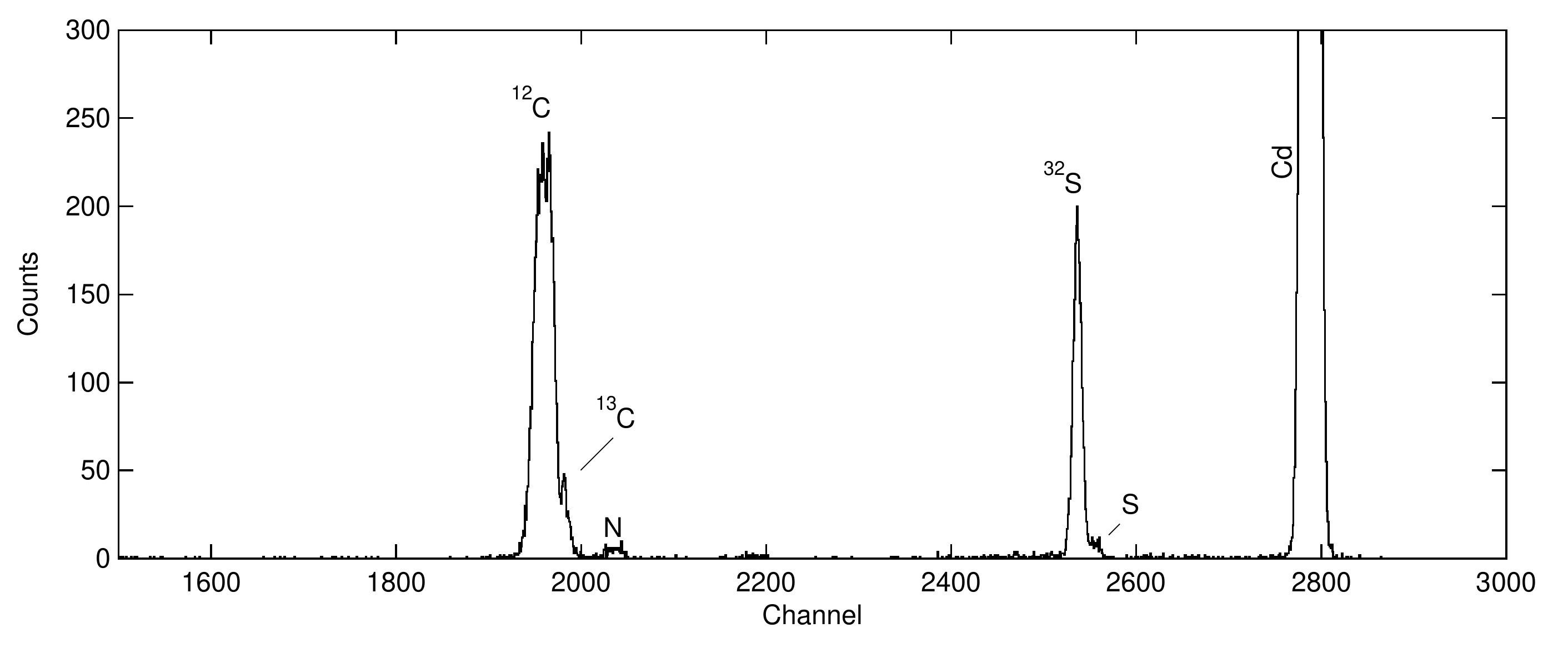}\\
\end{center}\vspace{-0.5cm}
\caption{\label{figure3}The momentum spectrum of $\alpha$-particles elastically scattered off of the CdS target at 30$^{\circ}$. The spectrum is obtained using the TUNL Enge split-pole spectrograph and its focal plane detector package. The small shoulder labelled by S refers to other stable isotopes of sulfur in the CdS target. The negligible nitrogen contamination does not show up on the RBS spectra and may be coming from vacuum contamination of the target chamber. No excited states produced from ($\alpha, p$) reactions on the nitrogen or $^{13}$C contamination of the CdS target were observed in the $^{32}$S($\alpha, p$) spectra. Due to kinematic broadening~\cite{Marshall:2018,Enge:1958,Spencer:1967,Enge:1979}, only the $\alpha$-particles scattered off of $^{32}$S are in focus.}
\end{figure*}
\begin{figure*}[ht]
\begin{center}
\includegraphics[width=0.6\textwidth]{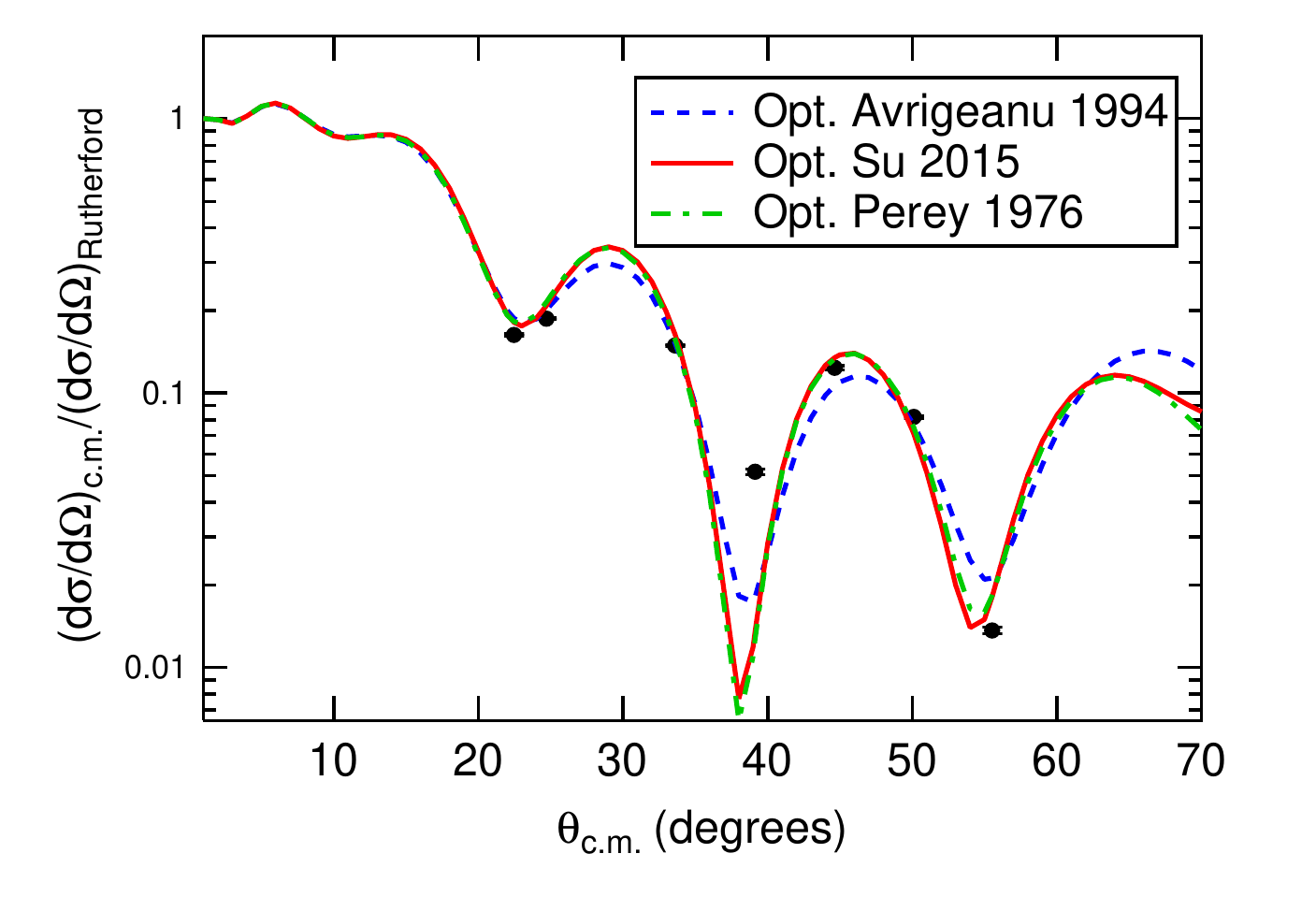}\\
\end{center}\vspace{-0.5cm}
\caption{\label{figure4}(Color online) The filled circles represent the angular distribution of the ratio of the center-of-mass differential to Rutherford $^{4}$He $+$ $^{32}$S elastic scattering cross section at 21 MeV. If not shown, the error bar is smaller than the point size. The curves are the theoretical DWBA calculations using FRESCO~\cite{Thompson:1988}. Each curve is computed via optimizing the parameters of a specific global optical potential model taken from Refs.~\cite{Avrigeanu:1994,Su:2015} and the $^{4}$He $+$ $^{32}$S potential of Ref.~\cite{Perey:1976} (see p.~83). The latter was obtained at 23.8 MeV. The global Su-2015~\cite{Su:2015} optimized models describes the data the best.}
\end{figure*}

Figure~\ref{figure3} shows the $\theta_{lab}$ $=$ 30$^{\circ}$ momentum spectrum of elastically scattered $^{4}$He beam particles off of the CdS target. A similar spectrum was obtained at each angle, where elastic scattering was measured. For each spectrum, the peak corresponding to the $^{32}$S content of the CdS target was fitted using a least-squares Gaussian fit function to obtain the peak's area. This was subsequently corrected run-by-run for the detector deadtime, which varied between 0.5\% and 5\%. The corrected areas were then converted (see Ref.~\cite{Setoodehnia:2011}) to their ratio to the center-of-mass Rutherford cross section for fitting. The uncertainty in these ratios arise from the statistical uncertainties in the peak areas. Finally, these experimental ratios were plotted vs.~the center-of-mass angle (see Fig.~\ref{figure4}).\par
Theoretical ($d\sigma/d\Omega$)$_{c.m.}$/($d\sigma/d\Omega$)$_{Rutherford}$ ratios were computed with FRESCO~\cite{Thompson:1988} using global $^{4}$He optical potentials of Ref.~\cite{Avrigeanu:1994} and Ref.~\cite{Su:2015}. Three Woods-Saxon optical potentials described in Ref.~\cite{Perey:1976} (p.~83) for $\alpha$-particles of 18.1 and 23.8 MeV were also used\footnote{Those potentials obtained at 18.1 MeV did not well describe the present elastic scattering data even after optimizing their potential parameters. Therefore, these optimized models are not presented in Fig.~\ref{figure4}.}. However, none of these

\noindent models described the data well enough. Therefore, a $\chi^{2}$ minimization code was developed so that the potential parameters of these models could be adjusted (one model at the time) to improve the agreement between theoretical and experimental cross sections. Except for $r_{0c}$, which is the Coulomb radius and was kept constant, all the other parameters of the previously mentioned optical potentials were allowed to be varied by a maximum of 30\%. This factor was chosen since smaller variations in potential parameters resulted in models that were not too different from the original ones. Larger variations or having no boundary on how much the parameters could be varied, on the other hand, resulted in optical potentials which had unreasonably large or small radii and diffuseness parameters. Comparing the outputs of FRESCO with the data for each set of the adjusted parameters, the program searched for a minimum $\chi^{2}$ using the \underline{gen}etic \underline{o}ptimization \underline{u}sing \underline{d}erivatives

\begin{table*}[ht]
\caption{\label{tab:2}Optical potential parameters for the present DWBA analysis of the $^{4}$He $+$ $^{32}$S elastic scattering (the first row) and the $^{32}$S($\alpha, p$)$^{35}$Cl reaction (the remaining rows) at $E_{\alpha}$ $=$ 21 MeV. For elastic scattering, the presented parameters are optimized by varying the original potential parameters (except $r_{0c}$) of Ref.~\cite{Su:2015} by 30\% and minimizing $\chi^{2}$ (see text). For the ($\alpha, p$) reaction, the potential depths were varied to reproduce the correct binding energies corresponding to each interaction.}
\begin{threeparttable}[e]
\centering
\setlength{\tabcolsep}{4pt} 
\begin{tabular}{cccccccccccc}
\toprule[1.0pt]\addlinespace[0.3mm] Interaction  & $V_{R}$ & $r_{R}$ & $a_{R}$ & $W_{D}$ & $r_{D}$ & $a_{D}$ & $V_{so}$ & $r_{so}$ & $a_{so}$ & $r_{0c}$ \\
 & (MeV) & (fm) & (fm) & (MeV) & (fm) & (fm) & (MeV) & (fm) & (fm) & (fm) \\ \hline\hline\addlinespace[0.6mm]
$\alpha$ $+$ $^{32}$S & 115.9 & 1.19 & 0.85 & 30.2 & 1.24 & 0.45 &  &  &  & 1.35 \\
\hspace{0.15cm}$p$ $+$ $^{35}$Cl\tnote{a} & $56.76\,-\,0.32\,E_{p}\tnote{b}$ & 1.17 & 0.75 & $12.14\,-\,0.25\,E_{p}$ & 1.34 & 0.53 & 6.2 & 1.01 & 0.75 & 1.25 \\
$p$ $+$ $^{32}$S\tnote{a} & $55.96\,-\,0.32\,E_{p}$ & 1.17 & 0.75 & $11.80\,-\,0.25\,E_{p}$ & 1.32 & 0.51 & 6.2 & 1.01 & 0.75 & 1.25 \\
\hspace{-0.15cm}$t$ $+$ $p$\tnote{c} & $V$ & 2.27 & 0.30 &  &  &  &  &  &  & 1.25 \\[0.2ex]
\hspace{0.1cm}$t$ $+$ $^{32}$S\tnote{d} & $V$ & 0.929 & 0.921 &  &  &  &  &  &  & 1.30 \\[0.2ex]
\bottomrule[1.0pt]
\end{tabular}
\begin{tablenotes}
\item[a] Adopted from the global potential of Ref.~\cite{Becchetti:1969}.
\item[b] $E_{p}$ is the laboratory kinetic energy (in MeV) of an outgoing proton from the $^{32}$S($\alpha, p$) reaction.
\item[c] Adopted from Ref.~\cite{Edwards:1973}.
\item[d] Adopted from Ref.~\cite{Das:2000}.
\end{tablenotes}
\end{threeparttable}
\end{table*}

\noindent (GENOUD) optimization function~\cite{Genoud}. GENOUD attempts to optimize the $\chi^{2}$ for a predetermined maximum number of \textit{generations}, after which the code outputs the best adjusted parameters for the optical potential used to initiate the calculations. Figure~\ref{figure4} also presents the theoretical ratios of differential to Rutherford cross sections for the $^{4}$He $+$ $^{32}$S elastic scattering at 21 MeV in the center-of-mass system using the aforementioned optimized potentials, the parameters of which are given in Table~\ref{tab:2}.\par
Lastly, a reduced $\chi^{2}$ analysis was performed for each set of the optimized optical potentials used. The best $\chi^{2}$/$\nu$ was obtained for the optimized global optical potential of Ref.~\cite{Su:2015} with its parameters adjusted (except $r_{0c}$) to describe the present $^{4}$He $+$ $^{32}$S elastic scattering data at $E_{\alpha}$ $=$ 21 MeV. The parameters of this optimized global optical potential (see Table~\ref{tab:2}) were used for the DWBA calculations for the $^{32}$S($\alpha, p$)$^{35}$Cl reaction which are presented in the next subsection.

\subsection{\label{Spin-Parities}Spin-parities of $^{35}$Cl excited states}

DWBA calculations were performed assuming one-step finite-range~\cite{Thompson:2009} triton transfer using FRESCO~\cite{Thompson:1988} in order to determine the $\ell$-transfers, and thus the spins and parities of the final $^{35}$Cl excited states. The distorted waves were computed for an optical interaction potential of the form~\cite{Perey:1976}:
\begin{eqnarray}
\setlength{\extrarowheight}{0.3cm}
U(r) & = & V_{c}(r_{0c}) - V_{R}f(r_{R},a_{R}) \nonumber \\
&  & - i\left(V_{I}f(r_{I},a_{I})\,-\,4W_{D}\,\dfrac{d}{dr_{D}}f(r_{D},a_{D})\right) \nonumber \\
&  & + \left(\dfrac{\hbar}{m_{\pi}c}\right)^{2}V_{so}\,\vec{l}\cdot\vec{\sigma}\,\dfrac{1}{r}\,\dfrac{d}{dr}\,f(r_{so},a_{so}),
\label{equation1}
\end{eqnarray}
\noindent where the first term is the Coulomb potential of a point charge with a uniformly charged sphere of radius $r_{0c}\,A^{1/3}$; the second and the third terms are the real and imaginary volume Woods-Saxon potentials, respectively; the next term is a derivative (surface) Woods-Saxon potential; and the last term is a spin-orbit potential, where $\ell$ is the orbital angular momentum, and $\vec{\sigma}$ $=$ 2$\vec{s}$ ($s$ is the spin angular momentum); $m_{\pi}$ is the pion mass; $c$ is the speed of light; $r_{0c}$ is the reduced charge radius ($R_{c} = r_{0c}\,A^{1/3}$); $r_{R}$, $r_{I/D}$ and $r_{so}$ are the reduced radii of the real, imaginary (index $I$ refers for the volume term, while index $D$ refers to the surface term) and the spin-orbit potentials, respectively; $a_{R}$, $a_{I/D}$ and $a_{so}$ are the diffuseness parameters of the real, imaginary and the spin-orbit potentials, respectively; and $V_{R}$, $V_{I}$ and $W_{D}$, and $V_{so}$ are the real, imaginary and spin-orbit depths of the potential wells, respectively. The function $f(r,a)$ is defined as~\cite{Perey:1976}:
\begin{equation}
f(r_{j},a_{j}) = \dfrac{1}{1\,+\,\exp\left(\dfrac{r\,-\,r_{j}A^{1/3}}{a_{j}}\right)},
\label{equation2}
\end{equation}
\noindent where $A$ is the atomic mass number; $r$ is the radius of the nucleus; and index $j$ refers to $R$ for real, $I$ for imaginary volume, $D$ for imaginary surface and $so$ for spin-orbit terms.\par
The parameters of the optical potentials used for the present $^{32}$S($\alpha, p$) DWBA analysis are given in Table~\ref{tab:2}.\par
The $\alpha$-particle's wave functions were computed from binding a triton (as a cluster) to a proton assuming a real Woods-Saxon potential, the parameters of which are given in Table~\ref{tab:2}. In addition to this potential, we also considered the widely used Reid soft core potential~\cite{Reid:1968} to derive the $\alpha$ wave functions from the $p$ -- $t$ interactions. The shapes of the angular distributions of protons' center-of-mass differential cross sections remained identical regardless of which of the two previously mentioned binding potentials were used for $p$ $+$ $t$ $\rightarrow$ $\alpha$. However, the magnitudes of the DWBA cross sections decreased by $\sim$ 6\% to 11\% (depending on the $J^{\pi}$ value) for the case of Reid soft core potential. This was not a source of concern for the present study because we are not interested in calculating the $^{32}$S($\alpha, p$) spectroscopic factors.\par
\begin{figure*}[ht]
 \begin{center}
  \subfloat{%
   \includegraphics[width=0.33\textwidth]{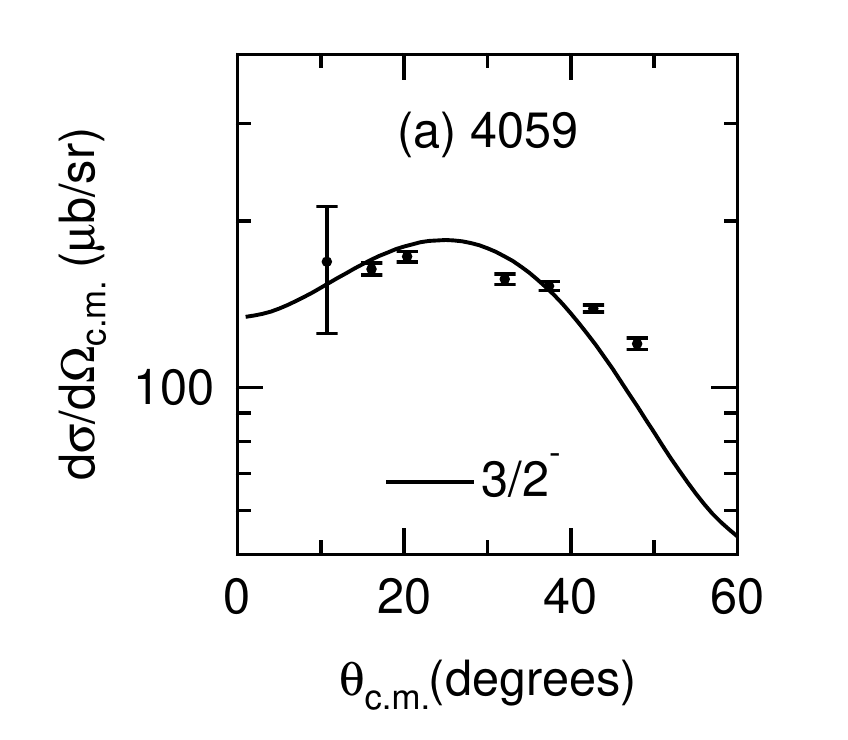}
   }
   \subfloat{%
   \includegraphics[width=0.33\textwidth]{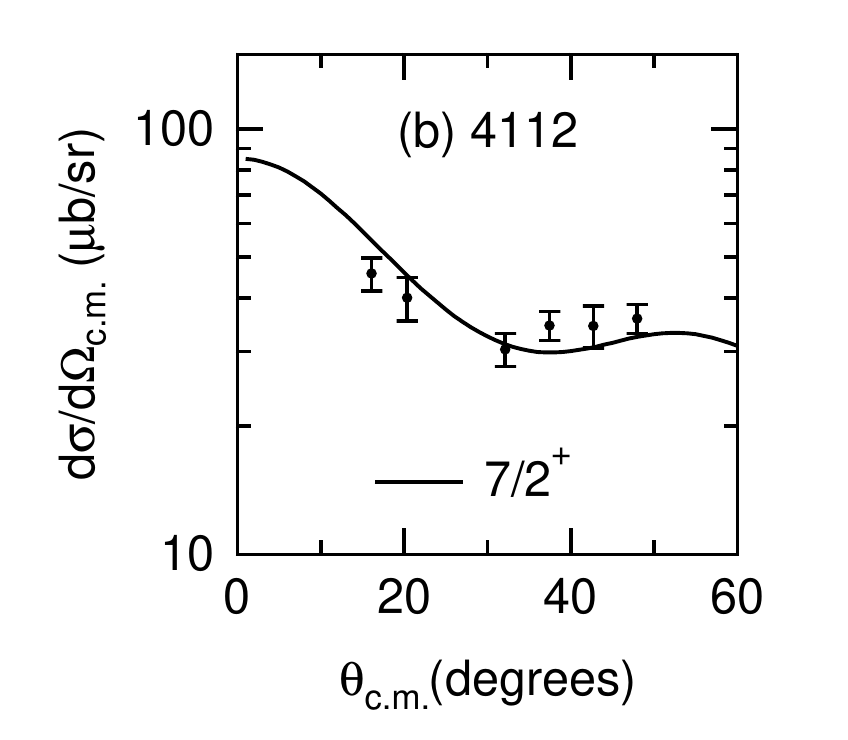}
   }
   \subfloat{%
   \includegraphics[width=0.33\textwidth]{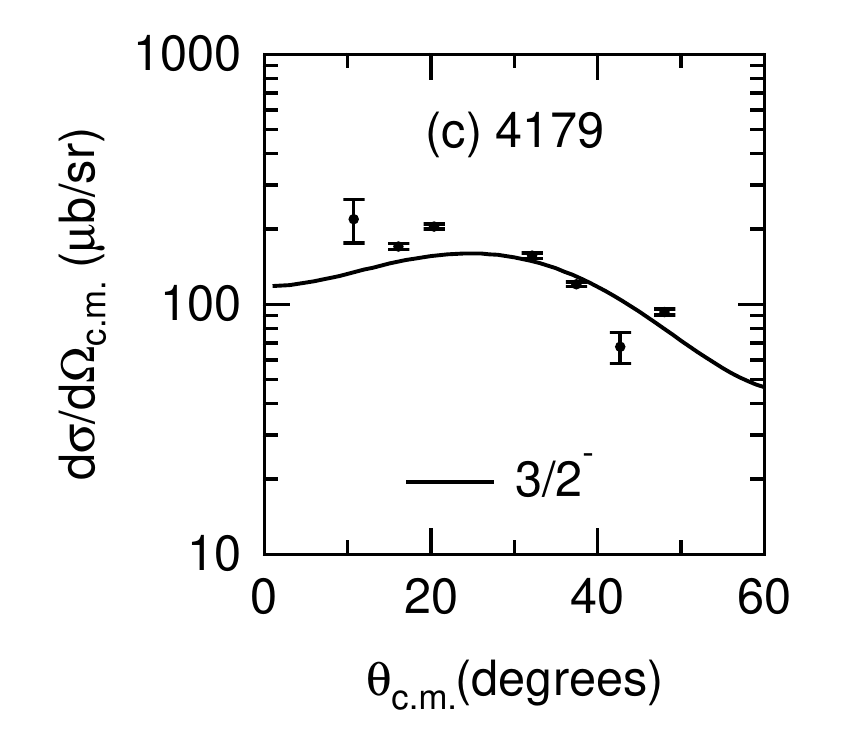}
   }\\
   \subfloat{%
   \includegraphics[width=0.33\textwidth]{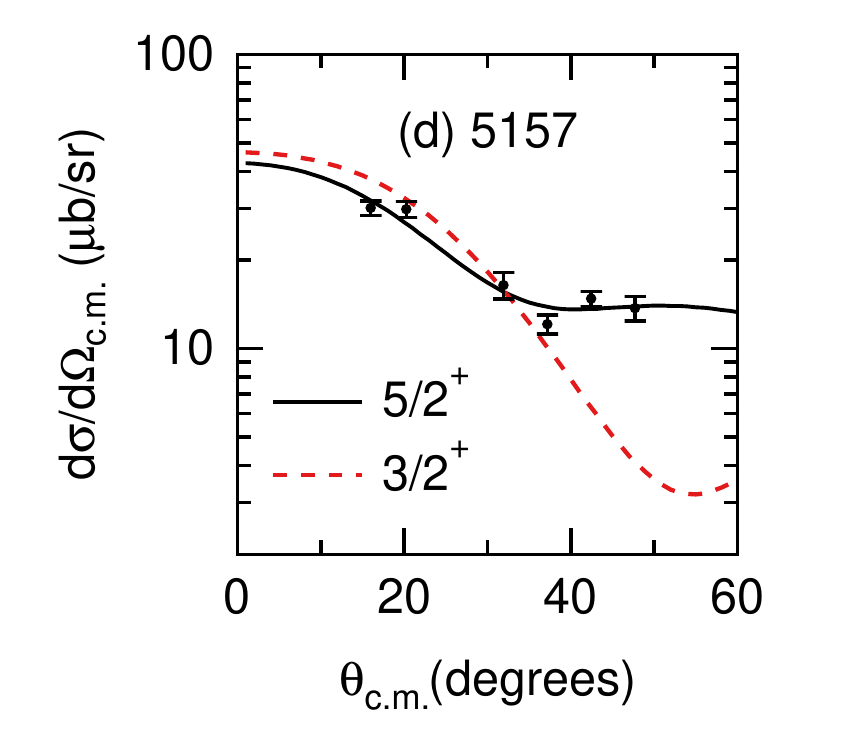}
   }
   \subfloat{%
   \includegraphics[width=0.33\textwidth]{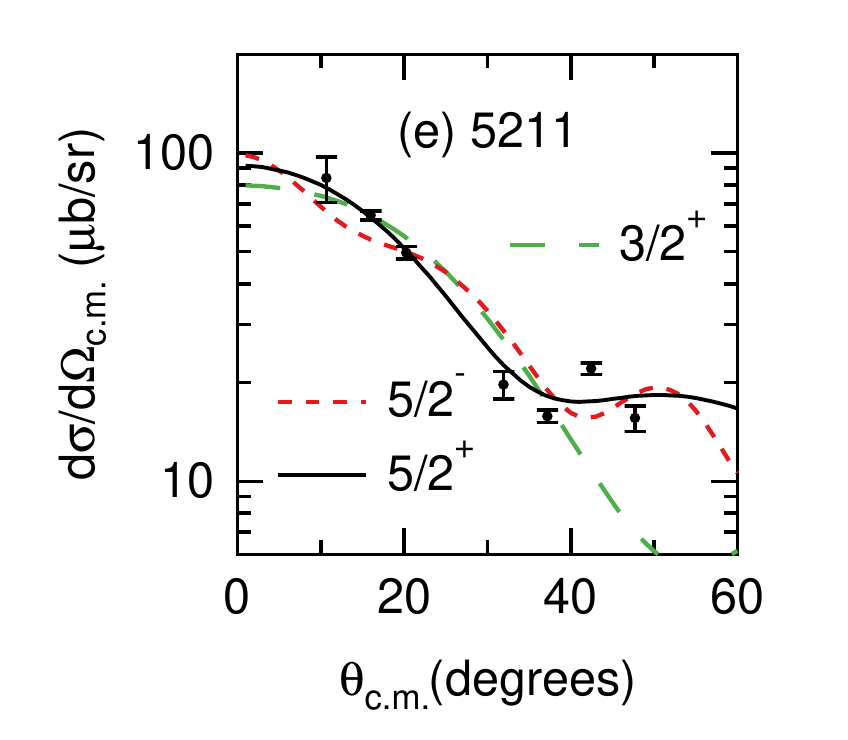}
   }
   \subfloat{%
   \includegraphics[width=0.33\textwidth]{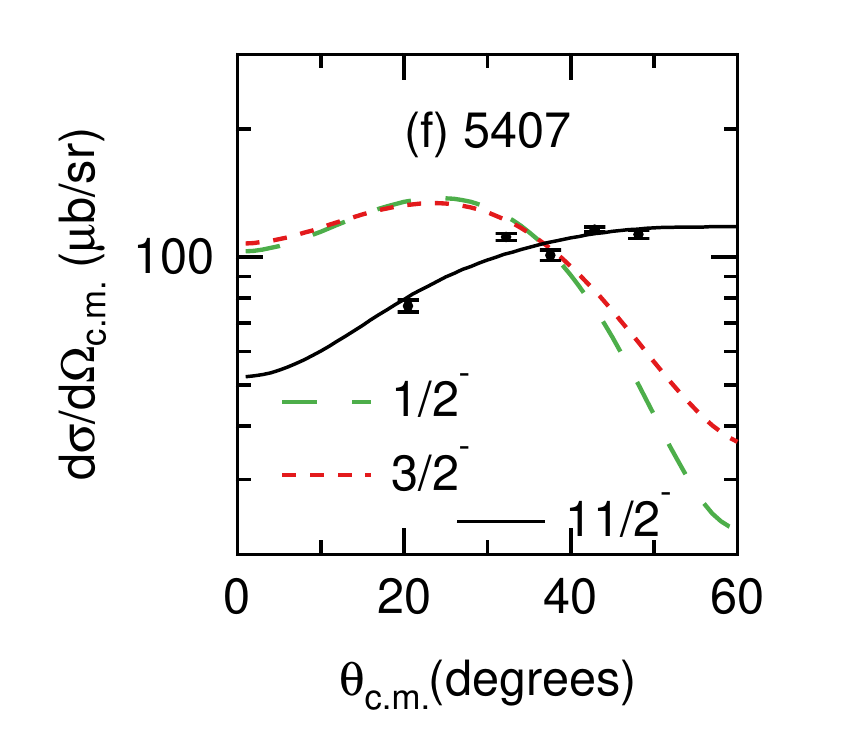}
   }\\
   \subfloat{%
   \includegraphics[width=0.33\textwidth]{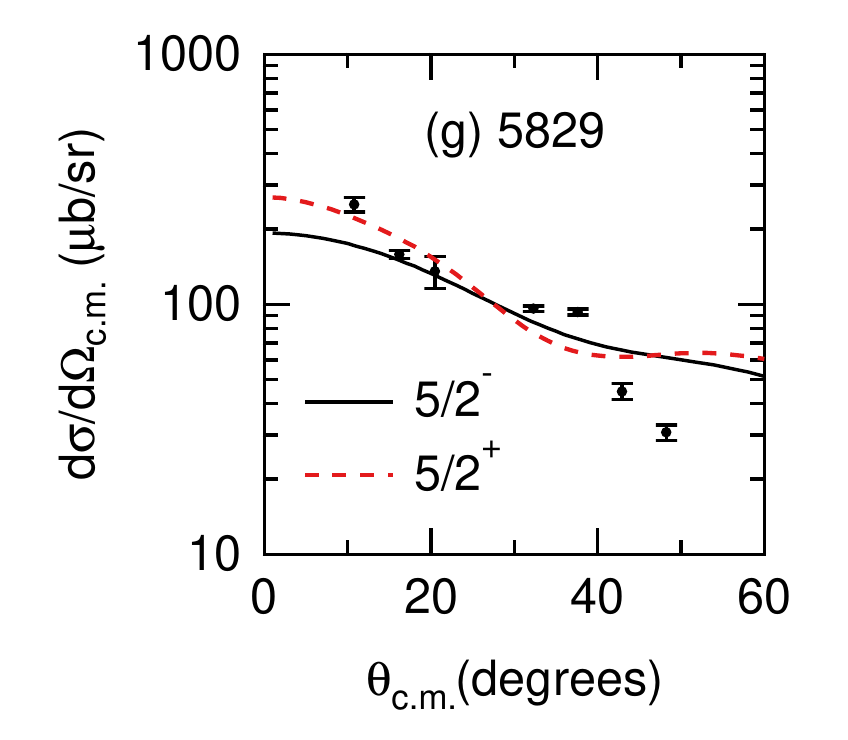}
   }
   \subfloat{%
   \includegraphics[width=0.33\textwidth]{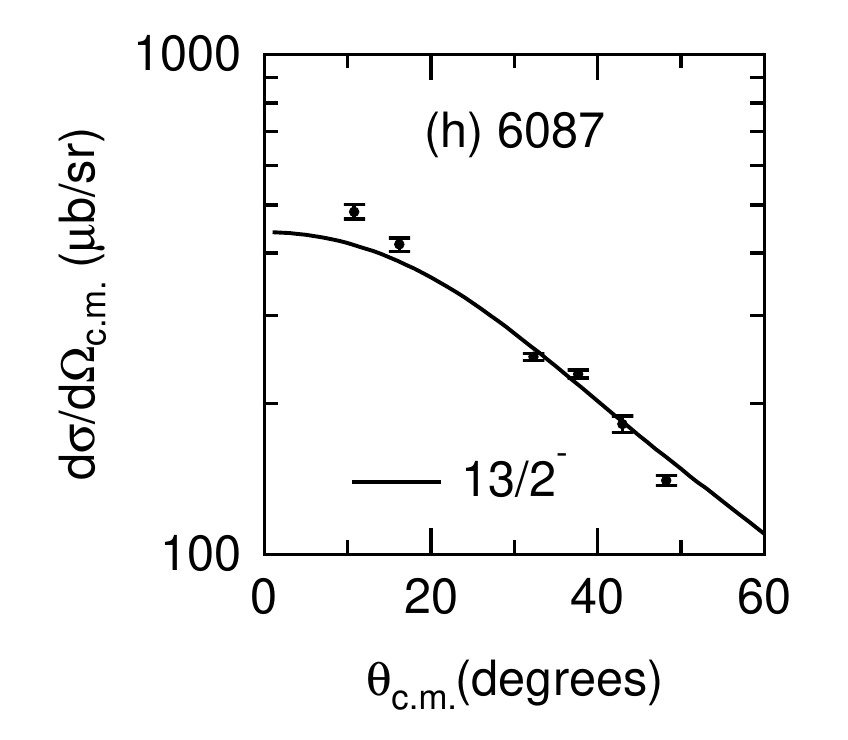}
   }
   \subfloat{%
   \includegraphics[width=0.33\textwidth]{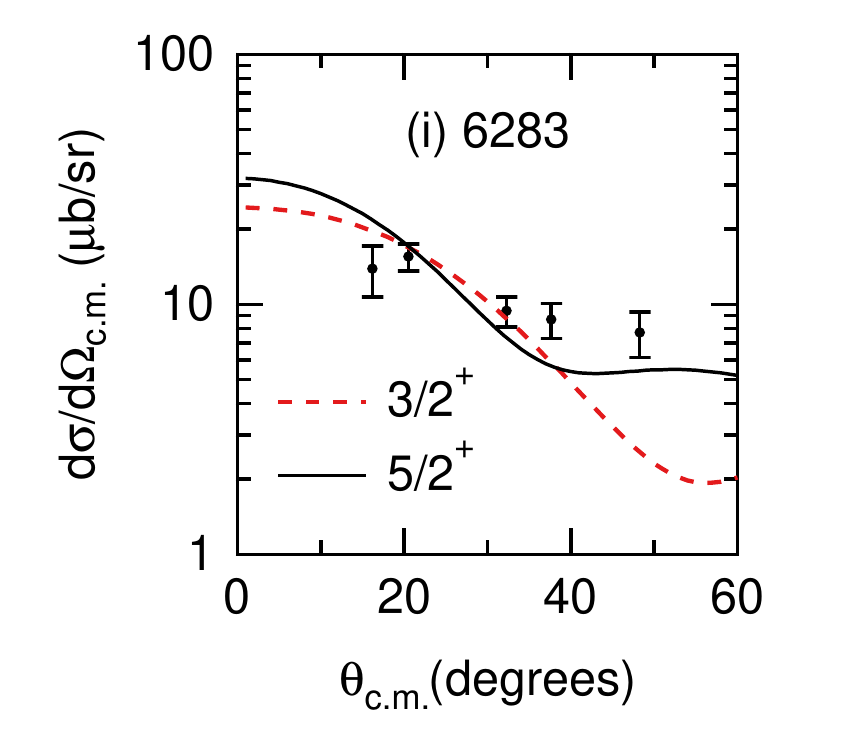}
   }\\
   \subfloat{%
   \includegraphics[width=0.33\textwidth]{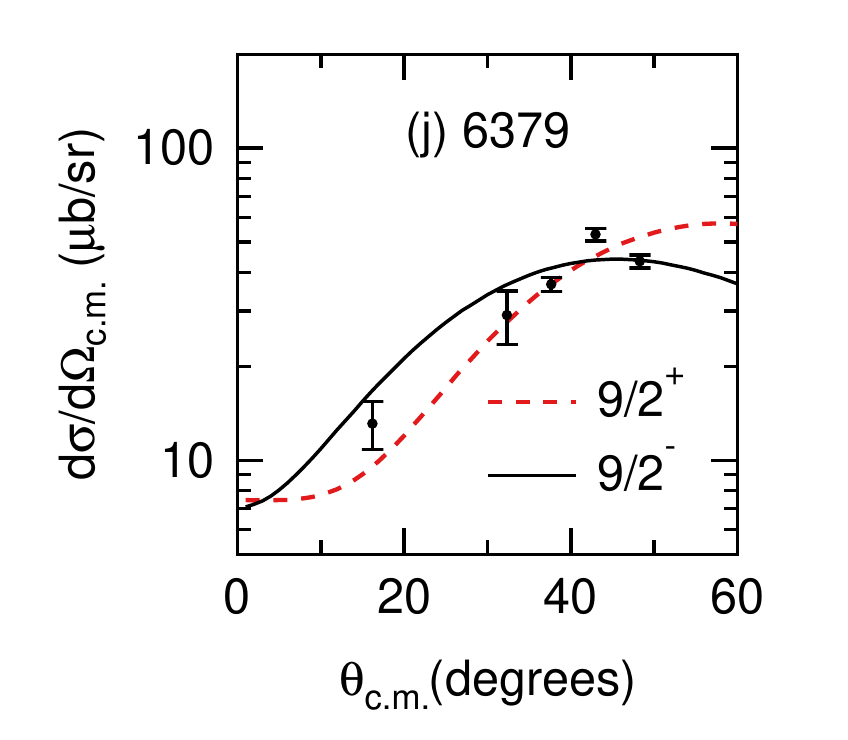}
   }
   \subfloat{%
   \includegraphics[width=0.33\textwidth]{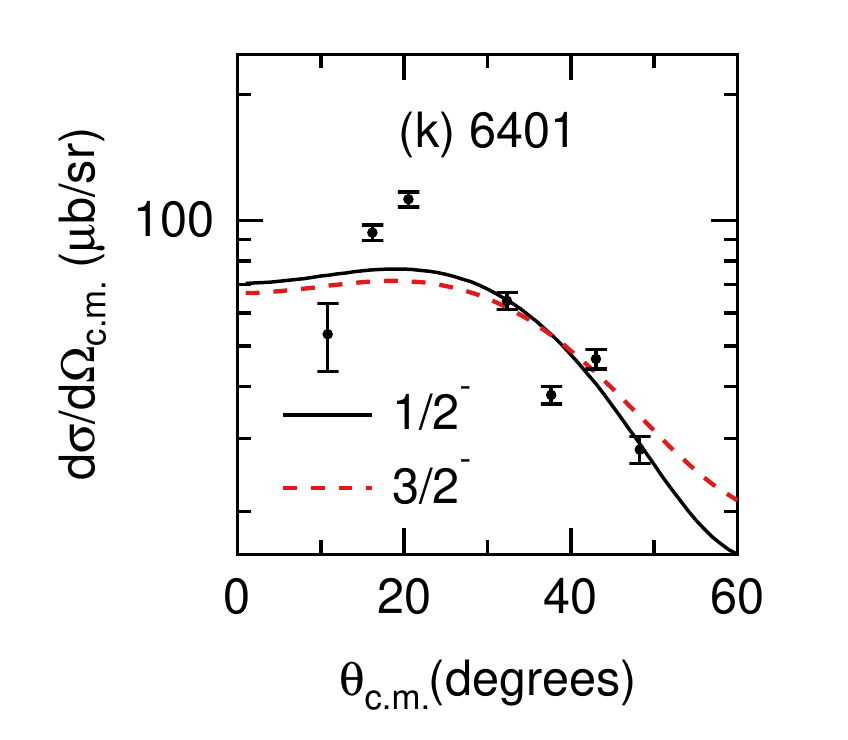}
   }
   \subfloat{%
   \includegraphics[width=0.33\textwidth]{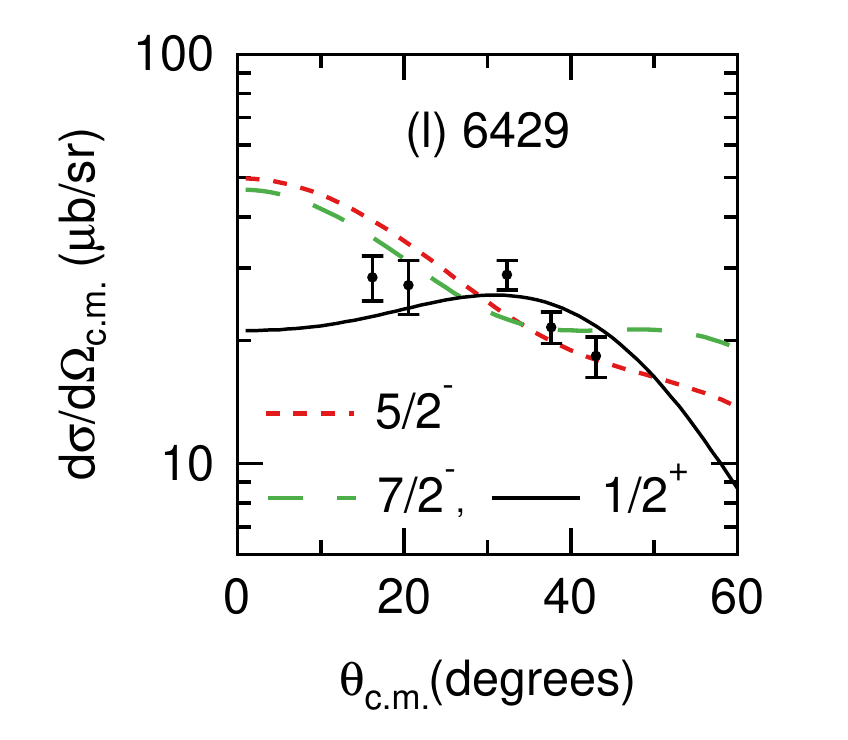}
   }
  \end{center}\vspace{-0.2cm}
\caption{\label{figure5}(Color online) Center-of-mass proton angular distributions of the $^{32}$S($\alpha, p$) reaction at 21 MeV (filled circles) compared with the DWBA curves (see legends) calculated using FRESCO~\cite{Thompson:1988}. If not shown, the error bar is smaller than the point size. The excited energies (in keV, rounded to the nearest integer) are given on the top of each panel.}
\end{figure*}
\begin{figure*}[ht]
 \begin{center}
  \subfloat{%
   \includegraphics[width=0.33\textwidth]{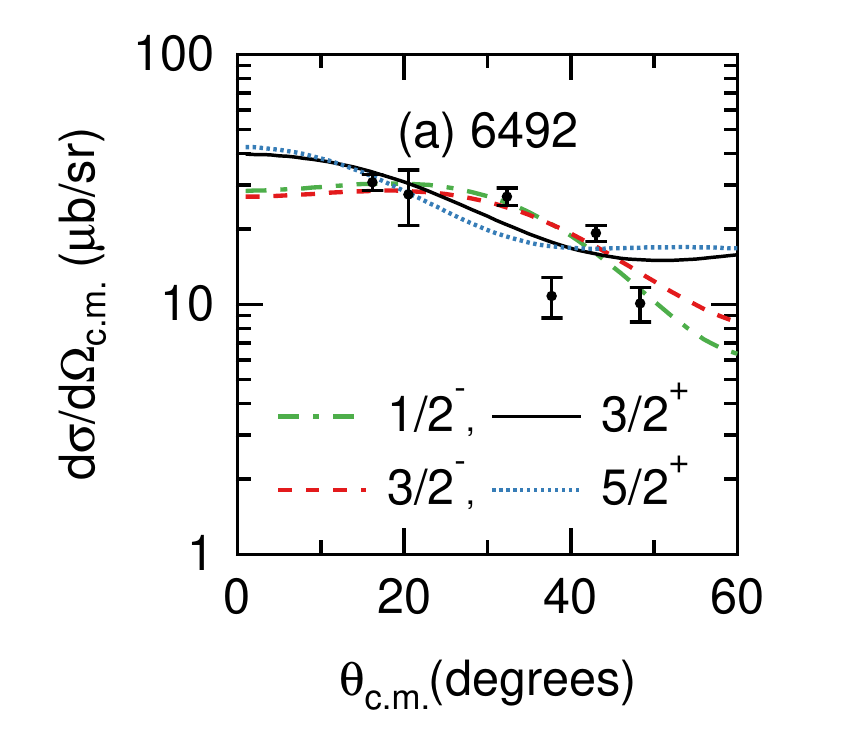}
   }
   \subfloat{%
   \includegraphics[width=0.33\textwidth]{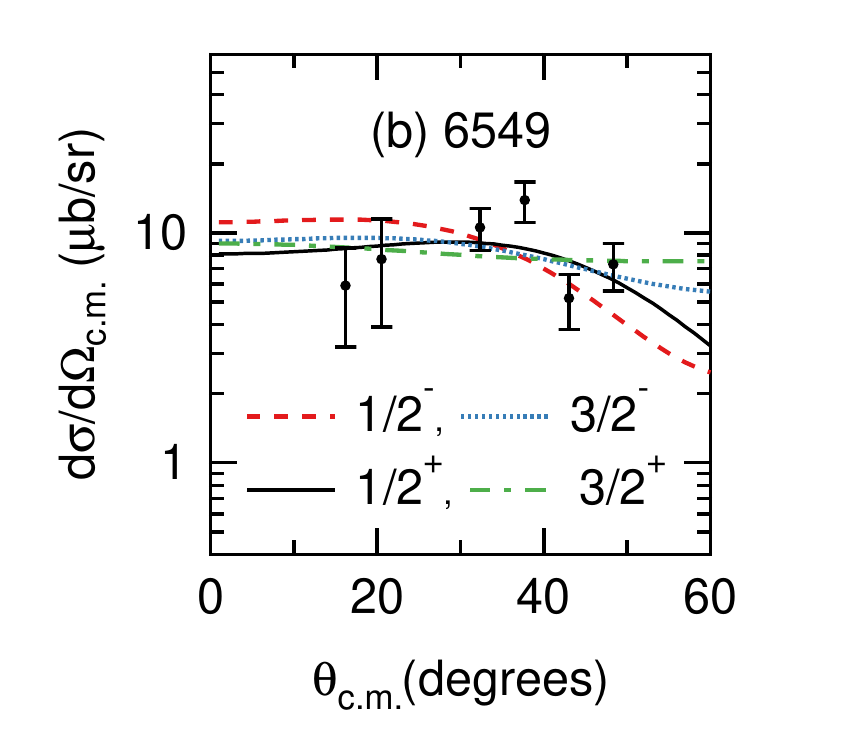}
   }
   \subfloat{%
   \includegraphics[width=0.33\textwidth]{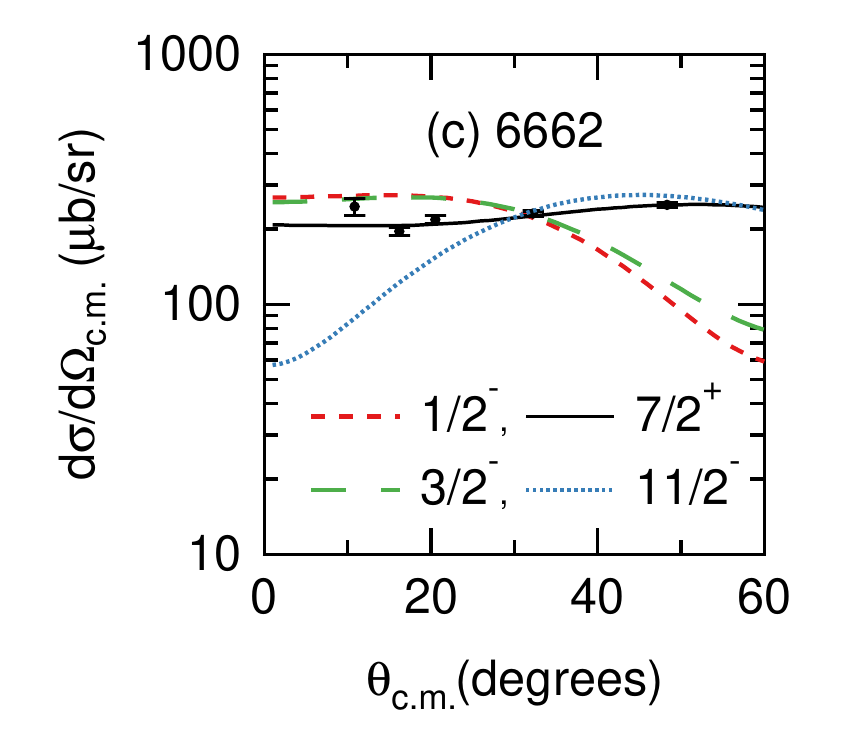}
   }\\
   \subfloat{%
   \includegraphics[width=0.33\textwidth]{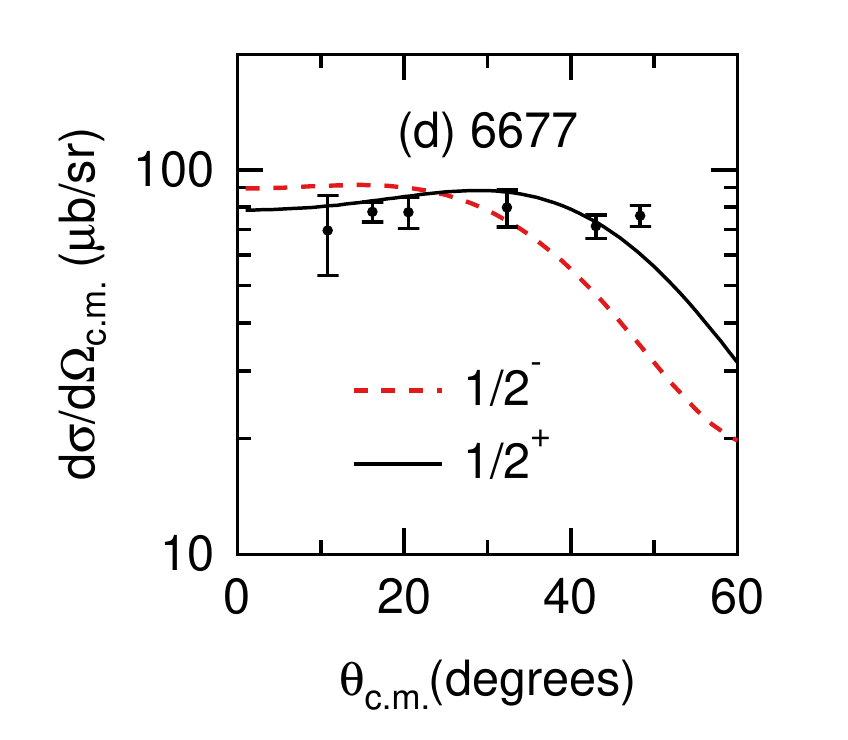}
   }
   \subfloat{%
   \includegraphics[width=0.33\textwidth]{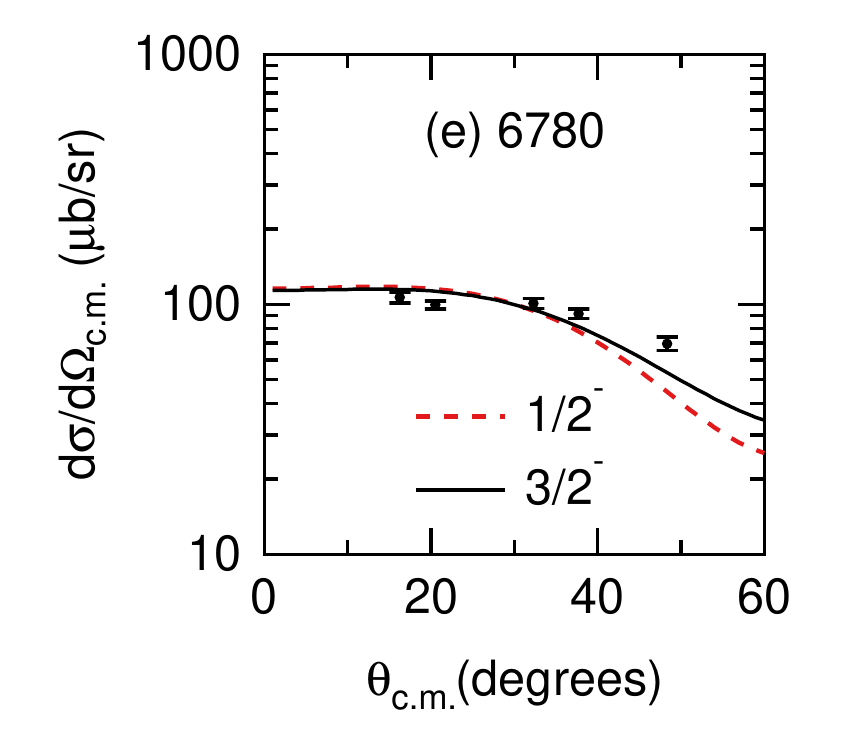}
   }
   \subfloat{%
   \includegraphics[width=0.33\textwidth]{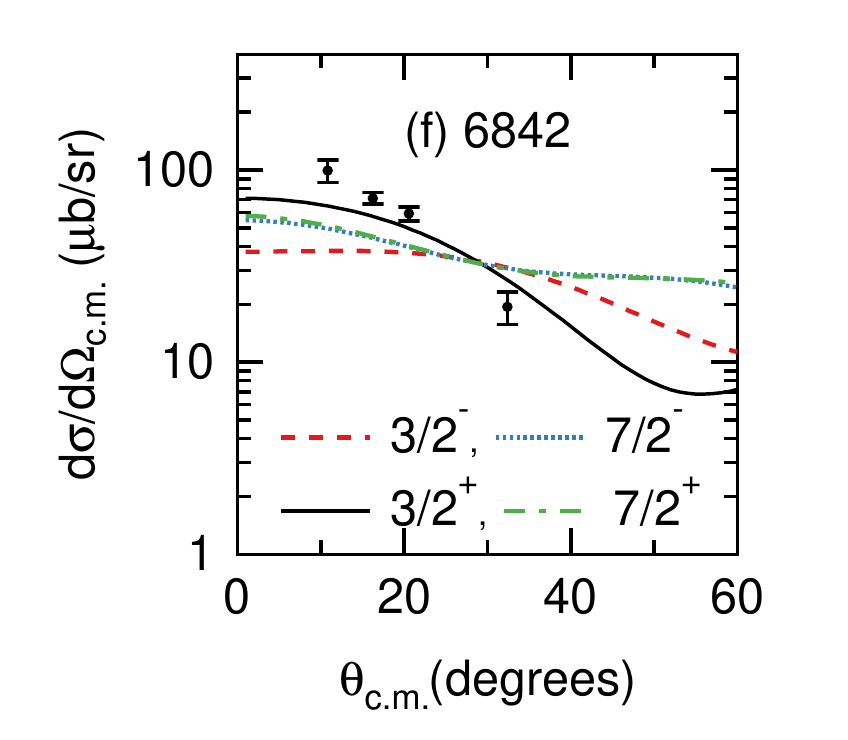}
   }\\
   \subfloat{%
   \includegraphics[width=0.33\textwidth]{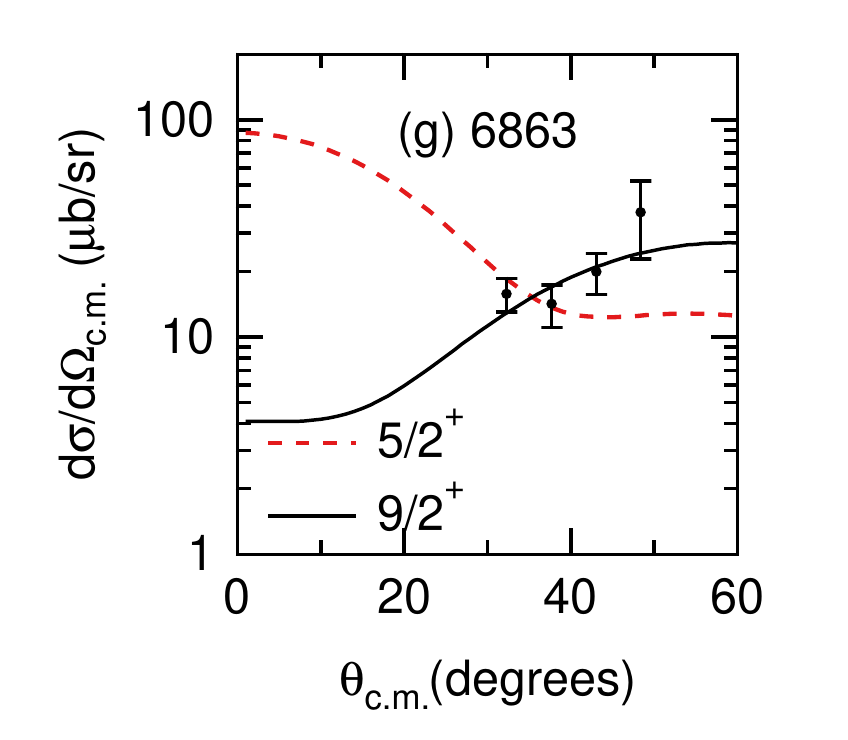}
   }
   \subfloat{%
   \includegraphics[width=0.33\textwidth]{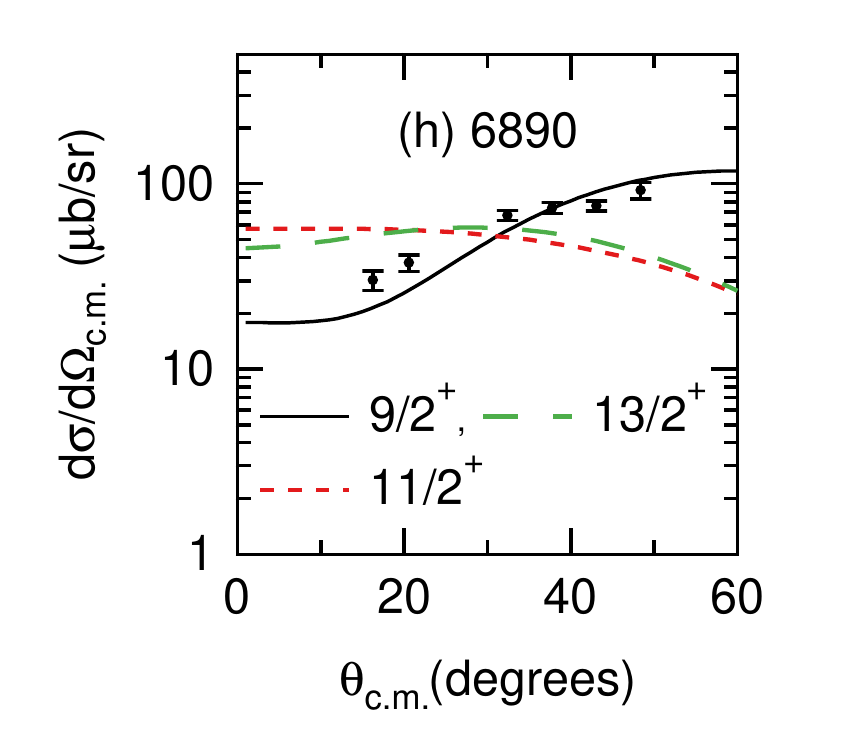}
   }
   \subfloat{%
   \includegraphics[width=0.33\textwidth]{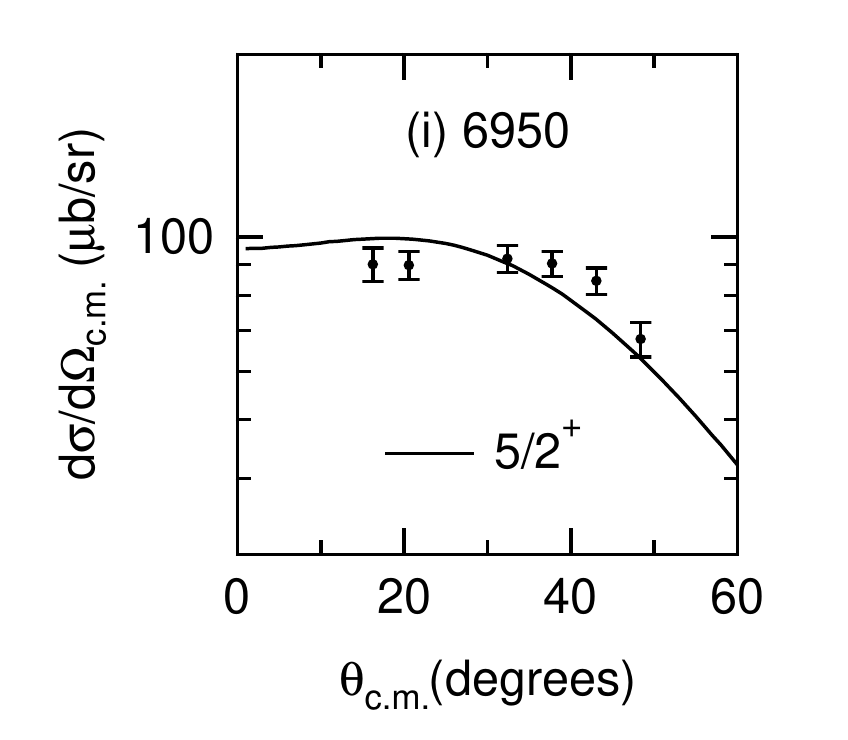}
   }
  \end{center}\vspace{-0.2cm}
\caption{\label{figure6}(Color online) Similar to Fig.~\ref{figure5}.}
\end{figure*}
\begin{figure*}[ht]
 \begin{center}
  \subfloat{%
   \includegraphics[width=0.33\textwidth]{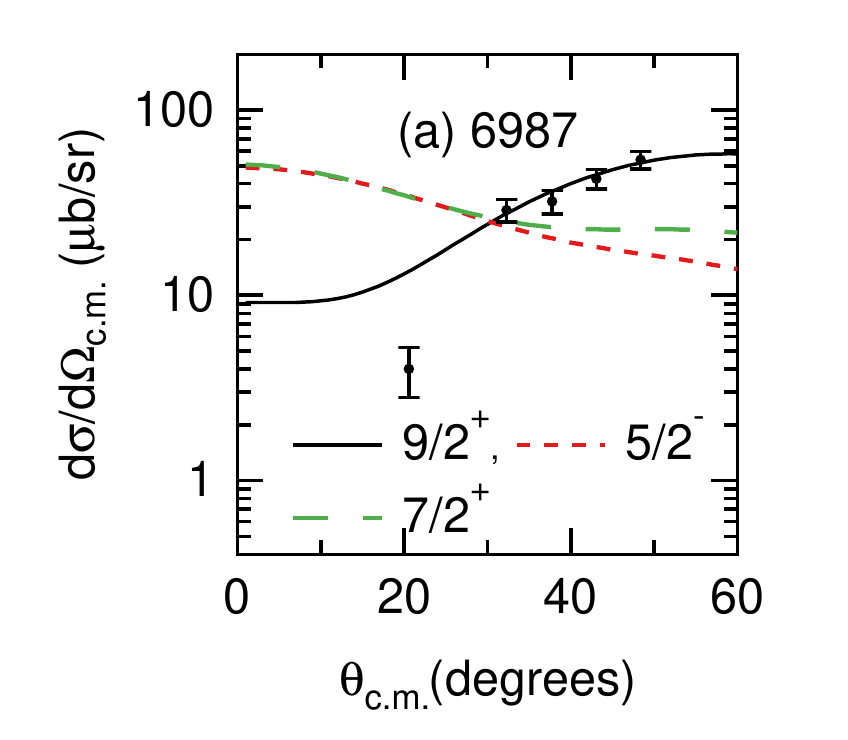}
   }
   \subfloat{%
   \includegraphics[width=0.33\textwidth]{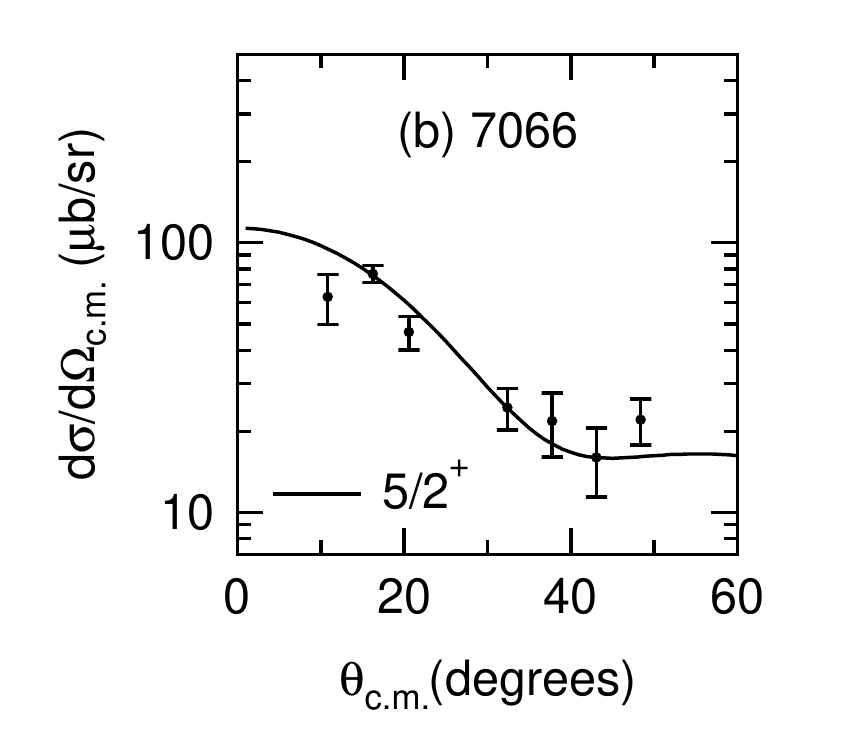}
   }
   \subfloat{%
   \includegraphics[width=0.33\textwidth]{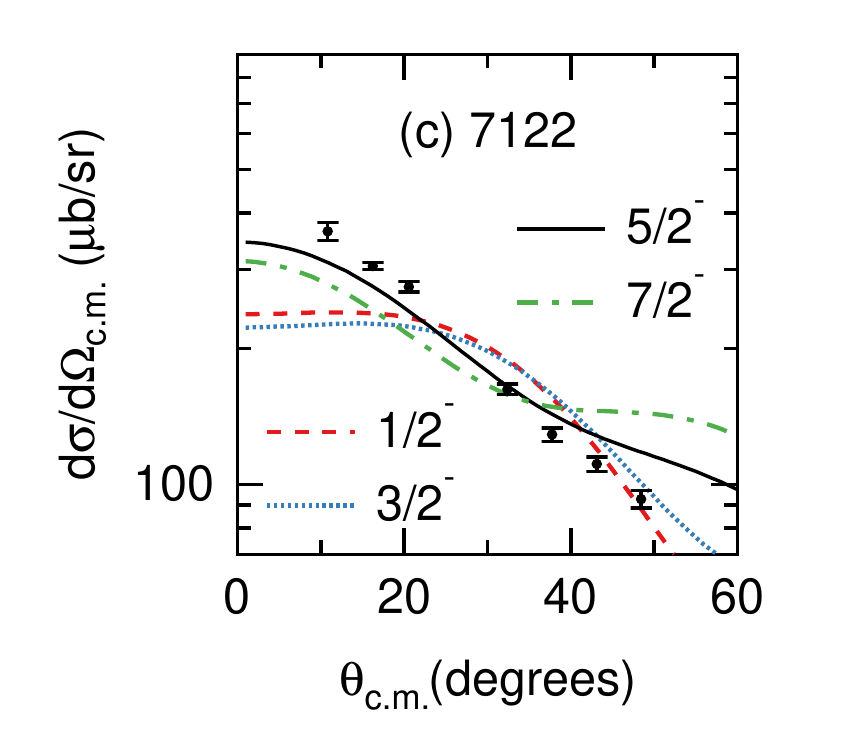}
   }\\
   \subfloat{%
   \includegraphics[width=0.33\textwidth]{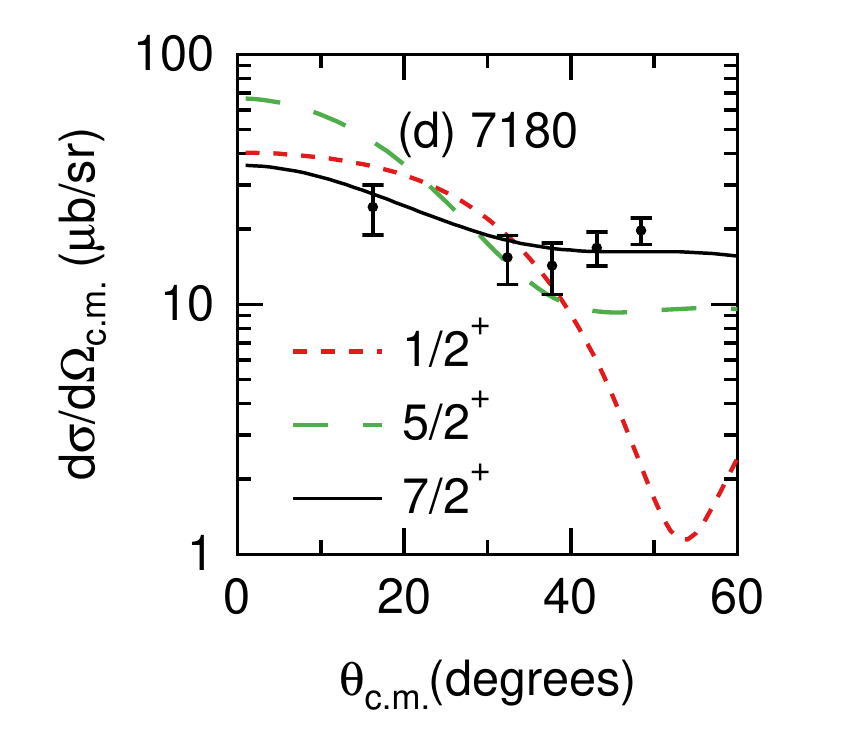}
   }
   \subfloat{%
   \includegraphics[width=0.33\textwidth]{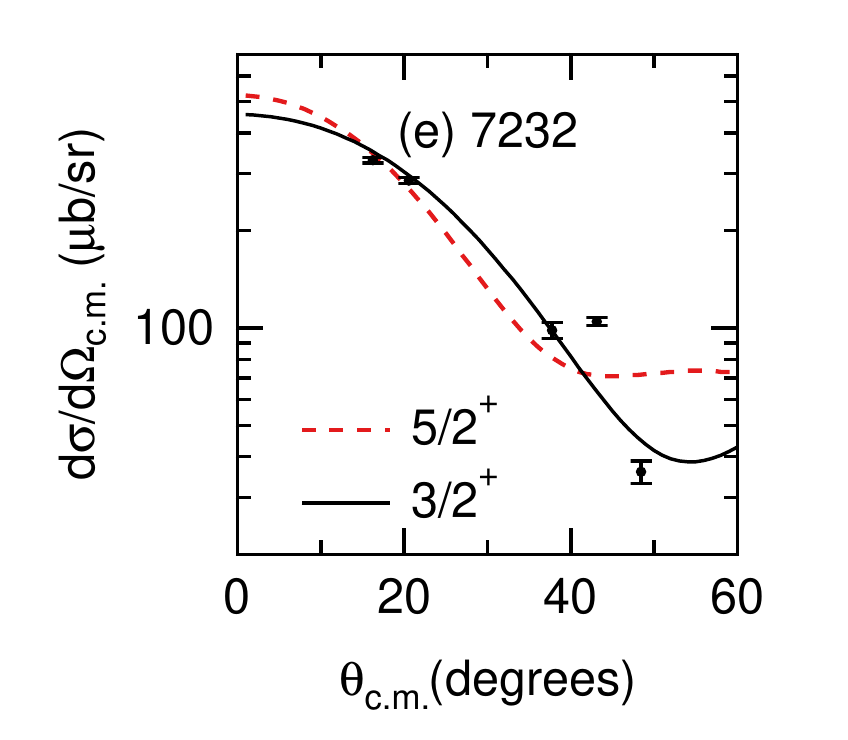}
   }
   \subfloat{%
   \includegraphics[width=0.33\textwidth]{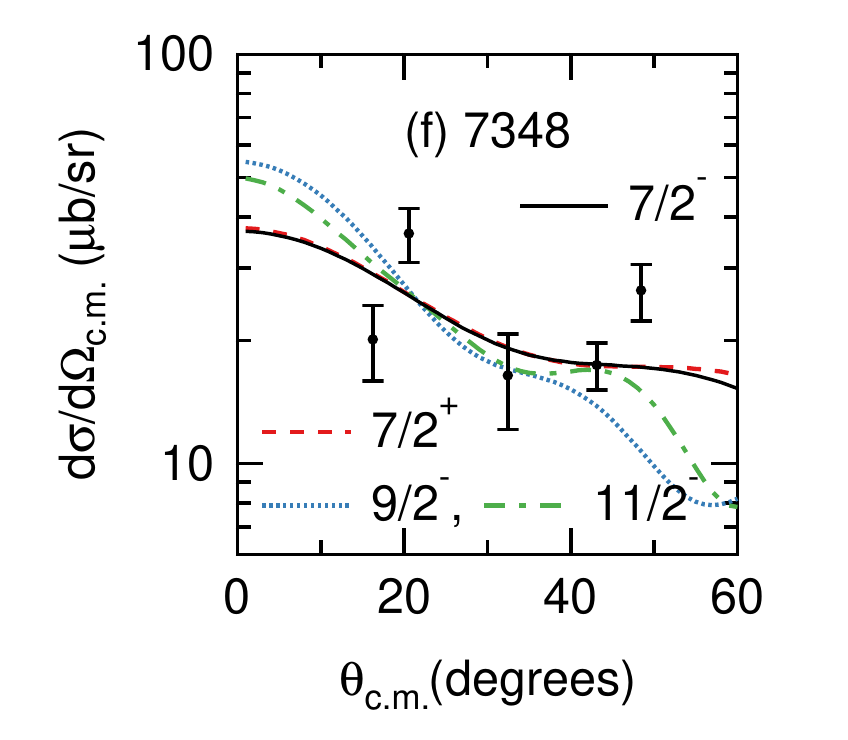}
   }
  \end{center}\vspace{-0.2cm}
\caption{\label{figure7}(Color online) Similar to Fig.~\ref{figure5}.}
\end{figure*}

The theoretical angular distribution curves were scaled to the center-of-mass experimental differential cross sections using linear fits with zero intercepts. If more than one $J^{\pi}$ values were consistent with the present data for a particular $^{35}$Cl state, the one with the minimum reduced $\chi^{2}$ of the fit is presented in Table~\ref{tab:1} as the present \textit{best} result for spin-parity of that state, and the corresponding DWBA curve is plotted in solid black lines in Figs.~\ref{figure5} to~\ref{figure7}.\par
Lastly, Refs.~\cite{Goss:1973,Das:2000} (and references therein) claim that the compound nuclear reaction mechanism probably contributes

\noindent to the three-nucleon transfer in the ($\alpha, p$) reactions. The contribution of the compound nucleus was not accounted for in the present study.\par
Figures~\ref{figure5} to~\ref{figure7} present the measured proton angular distributions for the $^{32}$S($\alpha, p$) reaction, as well as the theoretical DWBA fits for $^{35}$Cl excited states observed in this study at more than four angles. DWBA calculations were not performed for all the observed states. Figure~\ref{figure5} shows these calculations for a selection of the proton bound states in $^{35}$Cl, while Figs.~\ref{figure6} and~\ref{figure7} show such calculations for the proton resonances of interest for nova nucleosynthesis. The data at $\theta_{lab}$ $=$ 50$^{\circ}$ are not presented because the antimony sulfide target degraded during that measurement. Therefore, due to a sudden change (with an unknown amount) in the target thickness, reliable cross sections could not be extracted at $\theta_{lab}$ $=$ 50$^{\circ}$ so we excluded these data from the present DWBA analysis.\par
In what follows, we briefly compare the spin-parities derived in the present work with those obtained in previous measurements but only for the cases where the present assignments disagree with the previous ones, or if the present assignments are the only ones available. A matter of utmost concern that we should point out here is that the $\ell$-values in Table I of Ref.~\cite{Gillespie:2017} correspond to the $^{34}$S($^{3}$He, $d$) reaction, and there are at least 3 cases where the derived $\ell$-values are physically impossible: $\ell$ $=$ 1 for $E_{x}$($^{35}$Cl) $=$ 7066 keV with $J^{\pi}$ $=$ 5/2$^{+}$, $\ell$ $=$ 0 for $E_{x}$ $=$ 7273 keV with $J^{\pi}$ $=$ 1/2$^{-}$, and $\ell$ $=$ 2 for $E_{x}$ $=$ 7398 keV with $J^{\pi}$ $=$ 7/2$^{-}$. Considering the $J^{\pi}$ values of 1/2$^{+}$ and 1$^{+}$ for $^{3}$He and deuteron, respectively, these $\ell$ values violate the conservation of parity.\par
\textit{Selected proton bound states:} The present $J^{\pi}$ assignments of all these states (see Fig.~\ref{figure5}) except the 6429-keV state agree well with what is already known in the literature~\cite{Chen:2011,Gillespie:2017}.\par
\textit{The 6379.3-keV state:} The only information available in the literature regarding the spin and parity of this state is based on the measurement of Ref.~\cite{Gillespie:2017}, where a ($^{3}$He, $d$) angular momentum transfer of $\ell$ $=$ 2 or 3 was obtained. This implies that the $J^{\pi}$ assignments for this state would be 1/2$^{+}$, 3/2$^{+}$, 5/2$^{+}$ and 7/2$^{+}$ for $\ell$ $=$ 2; and 3/2$^{-}$, 5/2$^{-}$, 7/2$^{-}$ and 9/2$^{-}$ for $\ell$ $=$ 3. Out of these assignments, only $J^{\pi}$ $=$ 9/2$^{-}$ in addition to $J^{\pi}$ $=$ 9/2$^{+}$ describe our proton angular distributions well and are shown in panel (j) of Fig.~\ref{figure5}. $J^{\pi}$ $=$ 9/2$^{-}$ is the best fit to the present data with the minimum reduced $\chi^{2}$. We have tentatively assigned this state to be a 9/2$^{-}$ state.\par
\textit{The 6400.9-keV state:} No information regarding the spin and parity of this state is available in the literature~\cite{Chen:2011,Gillespie:2017}. The present proton angular distributions of this state seem to be best fitted with $J^{\pi}$ $=$ 1/2$^{-}$ and 3/2$^{-}$ assignments (see panel (k) of Fig.~\ref{figure5}). Out of these two assignments, $J^{\pi}$ $=$ 1/2$^{-}$ has a slightly better $\chi^{2}$/$\nu$ for the DWBA fit. We have therefore assigned for the first time a tentative $J^{\pi}$ $=$ (1/2$^{-}$) to this state.\par
\textit{The 6428.6-keV state:} This state was discovered in the measurement of Ref.~\cite{Gillespie:2017}, where tentative $J^{\pi}$ $=$ (5/2$^{-}$, 7/2$^{-}$) assignments were made based on a ($^{3}$He, $d$) $\ell$ $=$ 3 transfer. These assignments are fairly good fits to the present proton angular distribution data; however, $J^{\pi}$ $=$ 1/2$^{+}$ results in a slightly better reduced $\chi^{2}$ of the fit (see panel (l) of Fig.~\ref{figure5}). We have therefore assigned a tentative $J^{\pi}$ $=$ (1/2$^{+}$) to this state.\par
\textit{The 6491.8-keV state:} This state has tentative $J^{\pi}$ assignments of (1/2, 3/2 and 5/2$^{+}$) from Ref.~\cite{Chen:2011}. More recently, Ref.~\cite{Gillespie:2017} has obtained tentative $J^{\pi}$ assignments of (3/2$^{+}$, 5/2$^{+}$) based on a ($^{3}$He, $d$) $\ell$ $=$ 2 transfer. All these assignments are fairly good fits to the present proton angular distribution data (see panel (a) of Fig.~\ref{figure6}); however, the best fit with minimum reduced $\chi^{2}$ is obtained for $J^{\pi}$ $=$ 3/2$^{+}$.\par
\textit{The 6662.2-keV state:} There is a state in Ref.~\cite{Chen:2011} whose excitation energy is 6656(3) keV (from Ref.~\cite{Goss:1973}), which is consistent (within 2$\sigma$) with the one observed at 6662.2(19) in the present work. However, no $J^{\pi}$ assignment is available for this state from Refs.~\cite{Chen:2011,Goss:1973}. A recent measurement~\cite{Bisoi:2013} observed the $\gamma$-decay of a level at 6660 keV, to which they assigned $J^{\pi}$ $=$ 11/2$^{-}$. However, no uncertainty on the level energy is quoted in their result. In the measurement of Ref.~\cite{Gillespie:2017}, a new state was observed at 6643(2) keV with a pure ($^{3}$He, $d$) $\ell$ $=$ 1 transition. Therefore, they assigned that state to have tentative $J^{\pi}$ $=$ (1/2$^{-}$, 3/2$^{-}$) assignments. The present proton angular distribution data is inconsistent with $J^{\pi}$ $=$ 1/2$^{-}$, 3/2$^{-}$ and 11/2$^{-}$. However, a $J^{\pi}$ $=$ 7/2$^{+}$ assignment agrees well with the data (see panel (c) of Fig.~\ref{figure6}). Since it is unclear wether or not the 6656/6662-keV and 6643-keV states are the same, we have tentatively assigned $J^{\pi}$ $=$ (7/2$^{+}$) to the present 6662-keV state and have assumed it to be a different state from the one newly observed at 6643-keV~\cite{Gillespie:2017}.\par
\textit{The 6677-keV state:} A recent experiment was performed by Chipps {\textit et al}.~\cite{Chipps:2017}, where they observed via $^{37}$Cl($p, t$) a state at 6677(15) keV. This level was associated with the higher energy state of the previously described doublet at 6656(3)/6681(3) keV observed in Ref.~\cite{Goss:1973}. Reference~\cite{Chipps:2017} have made the first-ever constraint on the spin and parity assignment for the 6677-keV level and have considered it to be most likely of positive parity, with a spin assignment of (1/2, 3/2, 5/2, 7/2). Their best DWBA fit was achieved for the $J^{\pi}$ $=$ 1/2$^{+}$ assignment. Reference~\cite{Gillespie:2017} has also observed a state at 6674(2) keV with a tentative assignment of $J$ $=$ (1/2 -- 7/2) based on possible $\ell$ $=$ 1, 2 and 3 transfers for the $^{34}$S($^{3}$He, $d$) reaction. The present proton angular distribution data were fitted with these spins considering both negative and positive parities for each. Our data were best fitted with $J^{\pi}$ $=$ 1/2$^{+}$ (see panel (d) of Fig.~\ref{figure6}, where not all $J^{\pi}$ assignments are shown for clarity). This is consistent with the result of Ref.~\cite{Chipps:2017}. We have therefore, firmly assigned $J^{\pi}$ $=$ 1/2$^{+}$ to this level.\par
\textit{The 6863.1-keV state:} Reference~\cite{Gillespie:2017} has adopted a firm $J^{\pi}$ $=$ 5/2$^{+}$ assignment for this level based on the previous measurements, as well as their constraints on the angular momentum transfer for this state. Their $J^{\pi}$ $=$ 5/2$^{+}$ assignment is inconsistent with the present $J^{\pi}$ analysis (see panel (g) of Fig.~\ref{figure6}). The proton angular distribution data from our study is best fitted with a $J^{\pi}$ $=$ 9/2$^{+}$ assignment. However, since the present data are measured at relatively large angles, we have considered our assignment to be tentative. This excited state could correspond to the lowest energy proton resonance in $^{35}$Cl, whose strength is directly measured~\cite{Meyer:1976,Hubert:1972}, and we have used measured resonance strengths when available for the present $^{34}$S($p, \gamma$)$^{35}$Cl reaction rate calculation (see \S~\label{Rate}).\par
\textit{The 6890.4-keV and 6987-keV states:} No information is available in the literature~\cite{Chen:2011} regarding the spins and parities of these states. They were not observed in the measurement of Ref.~\cite{Gillespie:2017} either. In the present study, DWBA calculations were performed using $J^{\pi}$ $=$ 1/2 to 13/2 assignments with both negative and positive parities. The only $J^{\pi}$ assignment that resulted in a good fit for both cases was $J^{\pi}$ $=$ 9/2$^{+}$ (see panels (h) and (a) of Figs.~\ref{figure6} and~\ref{figure7}, respectively). We have therefore, tentatively assigned $J^{\pi}$ $=$ (9/2$^{+}$) to these levels for the first time.\par
\textit{The 7122.1-keV state:} The spin and parity of this state is unknown from the previous measurements~\cite{Chen:2011}, and it has not been observed in the measurement of Ref.~\cite{Gillespie:2017}. The DWBA calculation was performed for $J^{\pi}$ $=$ 1/2$^{\pm}$ to 11/2$^{\pm}$ assignments. For clarity, not all these assignments are presented in panel (c) of Fig.~\ref{figure7}. The $J^{\pi}$ $=$ 1/2$^{\pm}$, 3/2$^{\pm}$, 5/2$^{-}$ and 7/2$^{-}$ assignments fit the present proton angular distribution data of the 7122.1-keV state well, and $J^{\pi}$ $=$ 5/2$^{-}$ has the minimum reduced $\chi^{2}$ of the fit. We have therefore assigned the 7122.1-keV to have $J^{\pi}$ $=$ (5/2$^{-}$).\par
\textit{The 7180-keV state:} In the latest evaluation of $^{35}$Cl~\cite{Chen:2011}, there is a level at 7178.6(3) keV whose spin and parity is known to be 1/2$^{+}$. In Ref.~\cite{Chen:2011}, there are two other states in this vicinity: the 7170(10) keV level with $J^{\pi}$ $=$ (7/2 -- 17/2)$^{+}$, and the 7185.0(3) keV level with $J^{\pi}$ $=$ 5/2$^{+}$. In the measurement of Ref.~\cite{Gillespie:2017}, a state was observed at 7178(2) keV, for which a firm $J^{\pi}$ $=$ 1/2$^{+}$ assignment was established based on a ($^{3}$He, $d$) $\ell$ $=$ 2 transfer. Since the energy resolution of the present work is not sufficient to resolve the three states in this energy region, we have performed DWBA calculations with $J^{\pi}$ $=$ 1/2$^{+}$, 5/2$^{+}$ and 7/2$^{+}$ -- 17/2$^{+}$. Out of all these assignments, $J^{\pi}$ $=$ 7/2$^{+}$ is the best fit to the present proton angular distribution data (see panel (d) of Fig.~\ref{figure7}). We have thus assigned a tentative $J^{\pi}$ $=$ 7/2$^{+}$ to the present observed level and have paired it with the 7170(10) keV level of Ref.~\cite{Chen:2011}.\par
\textit{The 7231.6-keV state:} The most recent evaluation of $^{35}$Cl excited states~\cite{Chen:2011} lists two levels at 7225.5(3) keV with $J$ $=$ 5/2, and 7234.0(3) keV with $J^{\pi}$ $=$ 5/2$^{+}$. Reference~\cite{Gillespie:2017} has observed a state at 7227(2) keV, to which an orbital angular momentum transfer of $\ell$ $=$ 0 and 1 are assigned. Our energy resolution is not sufficient enough to resolve these states. The present proton angular distribution data are consistent with $J^{\pi}$ $=$ 3/2$^{+}$ and 5/2$^{+}$ (see panel (e) in Fig.~\ref{figure7}). However, the former yields a better reduced $\chi^{2}$ of the fit. We have therefore, considered a tentative $J^{\pi}$ $=$ (3/2$^{+}$) for the 7231.6-keV state. This is consistent with what Gillespie {\textit et al}.~\cite{Gillespie:2017} assigned to a level with an adopted energy of 7233.5(10) keV when calculating the reaction rate.\par
\textit{The 7347.9-keV state:} The spin and parity of this state is also not known from the previous measurements~\cite{Chen:2011}, and it remained unobserved in the measurement of Ref.~\cite{Gillespie:2017}. DWBA calculations were performed for $J^{\pi}$ $=$ 1/2$^{\pm}$ to 11/2$^{\pm}$ assignments (see panel (f) in Fig.~\ref{figure7}). The $J$ $=$ 1/2, 3/2, 5/2, 7/2, 9/2 and 11/2 assignments with negative parity, as well as $J^{\pi}$ $=$ 7/2$^{+}$ fit the data well, and $J^{\pi}$ $=$ 7/2$^{\pm}$ have almost identical and minimum reduced $\chi^{2}$ of the fit. Therefore, for the first time, we have tentatively assigned the 7347.9-keV state to have $J$ $=$ (7/2).\par

\begin{table*}[ht]
\caption{\label{tab:3}Resonance properties used to calculate the $^{34}$S($p, \gamma$)$^{35}$Cl reaction rate. The first column lists the adopted $^{35}$Cl excitation energies, which are determined using a weighted average of the present (taking into account the 2-keV systematic uncertainty in our excitation energies) and previous works~\cite{Gillespie:2017,Chen:2011}. The second column lists the $^{35}$Cl resonance energies based on $S_{p}$ $=$ 6370.81(4) keV~\cite{Wang:2017} and the adopted excitation energies. For $J^{\pi}$ values, see Table~\ref{tab:1} together with Table II of Ref.~\cite{Gillespie:2017} (and the following text).}
\centering
\setlength{\tabcolsep}{4pt} 
\begin{tabular}{ll@{\hspace*{0.1em}}lclc} 
\toprule[1.0pt]\addlinespace[0.6mm]
$E_{x}$ (keV) & $E_{r}^{c.m.}$ (keV) & J$^{\pi}$ & $\omega \gamma$ (eV) & ($2J + 1$)$C^{2}S$ & $\Gamma_{p}$ (eV) \\
(Adopted) &  &  & \cite{Meyer:1976,Hubert:1972} & \cite{Gillespie:2017} &  \\
\hline\hline\addlinespace[0.6mm]
\multirow{2}{*}{6427.5(20)} & \multirow{2}{*}{56.7(20) \hspace{0.3cm}\vast\{}    & (1/2$^{+}$)           &           & $<$ 2      & $<$ 1.47$\times$10$^{-18}$ \\
                            &                                                    & (5/2,7/2)$^{-}$       &           & 0.049      & 2.1(8)$\times$10$^{-24}$ \\ \addlinespace[0.6mm]
6471.5(31)                  & 100.7(31)                                          & (1/2,3/2)$^{-}$       &           & 0.034      & 9.6(38)$\times$10$^{-14}$ \\
\multirow{2}{*}{6491.9(6)}  & \multirow{2}{*}{121.1(6) \hspace{0.3cm}\vast\{}    & (3/2$^{+}$)	         &           & 0.072      & 2.3(9)$\times$10$^{-13}$ \\
                            &                                                    & (3/2$^{-}$)           &           & 0.080      & 3.9(42)$\times$10$^{-12}$ \\ \addlinespace[0.6mm]
\multirow{2}{*}{6546(2)}    & \multirow{2}{*}{175.2(20) \hspace{0.1cm}\vast\{}   & (1/2$^{+}$)	         &           & 0.004      & 5.7(27)$\times$10$^{-9}$ \\
                            &                                                    & (3/2$^{-}$)           &           & 0.0028     & 2.7(11)$\times$10$^{-10}$	\\ \addlinespace[0.6mm]
6643(2)                     & 272.2(20)                                          & (1/2,3/2)$^{-}$       &           & 0.0144     & 9.7(39)$\times$10$^{-6}$ \\
6659.1(28)                  & 288.3(28)                                          & (7/2$^{+}$)           &           & $<$ 8      & $<$ 6.53$\times$10$^{-8}$ \\
6677(3)                     & 306.2(30)                                          & 1/2$^{+}$	         &           & $<$ 2      & $<$ 2.53$\times$10$^{-4}$ \\
\multirow{2}{*}{6761(2)}    & \multirow{2}{*}{390.2(20) \hspace{0.1cm}\vast\{}   & (1/2$^{-}$,3/2$^{-}$) &           & 0.0032     & 1.5(6)$\times$10$^{-4}$ \\
                            &                                                    & (1/2$^{+}$)           &           & 0.0056     & 1.6(7)$\times$10$^{-3}$ \\ \addlinespace[0.6mm]
6780(2)                     & 409.2(20)                                          & (3/2$^{-}$)	         &           & 0.0084     & 6.9(28)$\times$10$^{-4}$ \\
\multirow{3}{*}{6800(4)}    & \multirow{3}{*}{429.2(40) \hspace{0.1cm}\vastp\{}  & (1/2$^{+}$)		     &           & $<$ 2      & $<$ 1.83 \\
                            &                                                    & (3/2$^{-}$)           &           & $<$ 4      & $<$ 5.81$\times$10$^{-1}$ \\
                            &                                                    & (3/2$^{+}$)           &           & $<$ 4      & $<$ 2.14$\times$10$^{-2}$ \\ \addlinespace[0.6mm]
6823(2)                     & 452.2(20)                                          & (1/2,3/2)$^{-}$	     &           & 0.006      & 3.2(14)$\times$10$^{-3}$ \\
6842(2)                     & 471.2(20)                                          & (3/2$^{+}$)           &           & 0.0216     & 3.5(14)$\times$10$^{-4}$\\
\multirow{2}{*}{6866.5(6)}  & \multirow{2}{*}{495.7(6) \hspace{0.3cm}\vast\{}    & (9/2$^{+}$)           &           & $<$ 10     & $<$ 6.97$\times$10$^{-5}$\\
                            &                                                    & 5/2$^{+}$             & $2.5(12) \times 10^{-2}$  &   & \\ \addlinespace[0.6mm]
6892(3)                     & 521.2(30)                                          & (9/2$^{+}$)	         &			 & $<$ 10	 & $<$ 2.14$\times$10$^{-4}$ \\
6949(4)                     & 578.2(40)                                          & 5/2$^{+}$		     &		     & $<$ 6	 & $<$ 6.16$\times$10$^{-1}$ \\
6987(4)                     & 616.2(40)                                          & (9/2$^{+}$)	         &		     & $<$ 10	 & $<$ 1.52$\times$10$^{-3}$ \\
7066.2(3)                   & 695.4(3)                                           & 5/2$^{+}$		     & $7.0(40) \times 10^{-2}$  &   &  \\
7103.3(3)                   & 732.5(3)                                           & 3/2$^{-}$		     & $2.3(12) \times 10^{-1}$  &	 &  \\
7122(2)                     & 751.2(20)                                          & (5/2$^{-}$)	         &			 & $<$ 6	 & $<$ 1.24$\times$10$^{+2}$ \\
\multirow{2}{*}{7178(2)}    & \multirow{2}{*}{807.2(20) \hspace{0.1cm}\vast\{}   & (7/2$^{+}$)           &		     & $<$ 8	 & $<$ 3.48$\times$10$^{-2}$ \\
                            &                                                    & 1/2$^{+}$             & $8.1(4) \times 10^{-1}$   & 	 & \\ \addlinespace[0.6mm]
7185.0(3)                   & 814.2(3)                                           & 5/2$^{+}$		     &		     & $<$ 6	 & $<$ 2.8$\times$10$^{+1}$ \\
7194(2)                     & 823.2(20)                                          & 1/2$^{-}$		     & $3.8(19) \times 10^{-1}$  &   & \\
7213.8(24)                  & 843.0(24)                                          & (1/2$^{+}$) 		     &			 & $<$ 2	 & $<$ 1.36$\times$10$^{+3}$ \\
7225.5(3)                   & 854.7(3)                                           & 5/2		             & $7.6(38) \times 10^{-2}$  & 	 & \\
7234.0(3)                   & 863.2(3)                                           & (3/2$^{+}$)	         & $5.2(10) \times 10^{-1}$  &	 & \\
7272.6(3)                   & 901.8(3)                                           & 1/2$^{-}$		     & $5.9(12) \times 10^{-1}$  &	 & \\
7347.9(27)                  & 977.1(27)                                          & (7/2$^{-}$)	         & 			 & $<$ 8	 & $<$ 4.01 \\
7362.0(3)                   & 991.2(3)                                           & 3/2$^{-}$		     & $8.5(17) \times 10^{-1}$  &	 & \\
7396.0(3)                   & 1025.2(3)                                          & 7/2$^{-}$		     & $1.9(10) \times 10^{-1}$  &	 & \\
\bottomrule[1.0pt]
\end{tabular}
\end{table*}

\section{\label{Rate}The $^{34}$S($p, \gamma$)$^{35}$\texorpdfstring{C\MakeLowercase{l}}{Cl} reaction rate}

Proton resonances dominating the $^{34}$S($p, \gamma$)$^{35}$Cl reaction rate over the nova temperature regime of 0.1 -- 0.4 GK are at energies of $E_{r}^{c.m.} = 122$ -- 556 keV. These correspond to the excitation energy range of 6493 keV $\lesssim$ $E_{x}$ $\lesssim$ 6927 keV in $^{35}$Cl ($Q$ $=$ 6370.81(4) keV~\cite{Wang:2017}).\par
The rate of the $^{34}$S($p, \gamma$)$^{35}$Cl reaction at a grid of temperatures, $T$, was calculated using the narrow resonance formalism and summing over each resonance, $i$:
\begin{equation}
\label{eq:rates-narrowrate}
N_{A}\langle\sigma v\rangle = N_{A}\left(\frac{2\pi}{\mu
kT}\right)^{3/2}\hbar^{2} \sum_{i} \omega\gamma_{i}\,e^{-E_{r,i}/kT},
\end{equation}
where $N_{A}$ is Avogadro's number, $\mu$ is the reduced mass of the reaction entrance channel, $k$ is the Boltzmann constant, $E_{r,i}$ are the center-of-mass resonance energies, and $\omega \gamma_{i}$ are the resonance strengths. For directly measured resonance strengths from Refs.~\cite{Meyer:1976,Hubert:1972} and summarized in Ref.~\cite{Gillespie:2017} (see the 4$\textsuperscript{th}$ column in Table~\ref{tab:3}), they enter directly into Eqn.~\ref{eq:rates-narrowrate}. Otherwise, they can be calculated using
\begin{equation}
\label{eq:rates-ResStrength}
\omega \gamma = \omega\,\frac{\Gamma_{p} \Gamma_{\gamma}}{\Gamma}.
\end{equation}

\noindent Here, $\omega$ is the spin factor, and $\Gamma_{p}$, $\Gamma_{\gamma}$, and $\Gamma$ are the proton, $\gamma$-ray, and total widths, respectively. The proton partial widths can be inferred from the spectroscopic factors ($C^{2}S$) obtained in Ref.~\cite{Gillespie:2017} (see the 5$\textsuperscript{th}$ column in Table~\ref{tab:3}) using a model uncertainty of 40\%:
\begin{equation}
\label{eq:Gammap}
\Gamma_{p} = \frac{2 \hbar^{2}}{\mu R^{2}}\,C^{2} S\, P_{\ell} \theta_{sp}^{2}\,.
\end{equation}
$P_{\ell}$ is the penetrability of the Coulomb and angular momentum barriers at the resonance energy, and $\theta_{sp}^{2}$ is the single-particle reduced width, which we estimated from the findings of Ref.~\cite{Iliadis:2007}.\par
In Ref.~\cite{Gillespie:2017}, it was assumed that $\Gamma_{p} \ll \Gamma_{\gamma}$, which implies $\omega \gamma \approx \Gamma_{p}$. By considering the average known lifetimes of excited states close to the excitation energies of interest, we estimate that the $\gamma$-ray partial widths are on the order of $\Gamma_{\gamma}$ $\approx$

\begin{table*}[ht]
\caption{\label{tab:pgRate} Monte Carlo reaction rates for the $^{34}$S($p, \gamma$)$^{35}$Cl reaction. Shown are the low, median, and high rates, corresponding to the 16$\textsuperscript{th}$, 50$\textsuperscript{th}$, and 84$\textsuperscript{th}$ percentiles of the Monte Carlo probability density distributions. Also shown are the parameters ($\mu$ and $\sigma$) of the lognormal approximation to the actual Monte Carlo probability density, as well as the Anderson-Darling statistic (A-D). See Ref.~\cite{Longland:2010} for details.}
  \sisetup{
    table-alignment=center,
    retain-zero-exponent=true,
    input-symbols = {()},
    explicit-sign,
    table-space-text-pre = (,
    table-space-text-post = ),
    table-align-text-pre = false,
    detect-all = true
  }
  \resizebox{!}{90mm}{
    \begin{tabular}{
      S
      S[table-format=4.2e4]
      S[table-format=3.2e4]
      S[table-format=3.2e4]
      S[table-format=3.3e2]
      S[table-format=3.2e2]
      S[table-format=3.2e2]
      }
      \toprule \toprule
      {T (GK)} & {Low rate} & {Median rate} & {High rate} & {lognormal $\mu$} & {lognormal $\sigma$} & {A-D}    \\ \midrule
0.010 &  6.96e-45 &  2.44e-44  &       7.94e-44 &  -1.005e+02  &       1.21e+00  &  2.40e+00  \\
0.011 &  2.91e-42 &  8.38e-42  &       2.22e-41 &  -9.461e+01  &       1.02e+00  &  3.17e+00  \\
0.012 &  4.43e-40 &  1.07e-39  &       2.44e-39 &  -8.976e+01  &       8.55e-01  &  3.95e+00  \\
0.013 &  3.05e-38 &  6.47e-38  &       1.29e-37 &  -8.566e+01  &       7.25e-01  &  4.52e+00  \\
0.014 &  1.13e-36 &  2.15e-36  &       3.88e-36 &  -8.215e+01  &       6.20e-01  &  4.72e+00  \\
0.015 &  2.54e-35 &  4.42e-35  &       7.41e-35 &  -7.912e+01  &       5.38e-01  &  4.31e+00  \\
0.016 &  3.81e-34 &  6.20e-34  &       9.81e-34 &  -7.647e+01  &       4.75e-01  &  3.27e+00  \\
0.018 &  3.39e-32 &  5.06e-32  &       7.48e-32 &  -7.207e+01  &       3.98e-01  &  6.84e-01  \\
0.020 &  1.36e-30 &  1.97e-30  &       2.89e-30 &  -6.839e+01  &       3.83e-01  &  9.53e-01  \\
0.025 &  9.22e-27 &  1.89e-26  &       3.87e-26 &  -5.923e+01  &       7.16e-01  &  3.16e-01  \\
0.030 &  1.91e-23 &  3.35e-23  &       5.88e-23 &  -5.176e+01  &       5.61e-01  &  1.97e+00  \\
0.040 &  2.71e-19 &  4.03e-19  &       5.95e-19 &  -4.236e+01  &       3.97e-01  &  9.86e-01  \\
0.050 &  7.52e-17 &  1.15e-16  &       1.83e-16 &  -3.668e+01  &       4.59e-01  &  6.73e+00  \\
0.060 &  3.13e-15 &  5.19e-15  &       9.81e-15 &  -3.283e+01  &       5.80e-01  &  4.05e+01  \\
0.070 &  5.26e-14 &  8.55e-14  &       1.82e-13 &  -2.998e+01  &       6.33e-01  &  1.03e+02  \\
0.080 &  5.73e-13 &  8.69e-13  &       1.71e-12 &  -2.766e+01  &       5.84e-01  &  1.35e+02  \\
0.090 &  4.31e-12 &  6.66e-12  &       1.12e-11 &  -2.569e+01  &       5.26e-01  &  4.39e+01  \\
0.100 &  2.20e-11 &  3.89e-11  &       6.38e-11 &  -2.399e+01  &       5.40e-01  &  3.71e+00  \\
0.110 &  8.99e-11 &  1.77e-10  &       3.05e-10 &  -2.249e+01  &       5.79e-01  &  2.23e+01  \\
0.120 &  3.58e-10 &  6.93e-10  &       1.26e-09 &  -2.111e+01  &       5.79e-01  &  4.19e+01  \\
0.130 &  1.43e-09 &  2.54e-09  &       4.51e-09 &  -1.979e+01  &       5.33e-01  &  4.28e+01  \\
0.140 &  5.45e-09 &  8.93e-09  &       1.48e-08 &  -1.853e+01  &       4.71e-01  &  2.14e+01  \\
0.150 &  1.89e-08 &  2.95e-08  &       4.49e-08 &  -1.735e+01  &       4.17e-01  &  7.28e+00  \\
0.160 &  5.87e-08 &  8.82e-08  &       1.28e-07 &  -1.626e+01  &       3.82e-01  &  3.82e+00  \\
0.180 &  4.21e-07 &  6.00e-07  &       8.38e-07 &  -1.434e+01  &       3.48e-01  &  1.57e+00  \\
0.200 &  2.21e-06 &  3.08e-06  &       4.23e-06 &  -1.269e+01  &       3.32e-01  &  1.44e+00  \\
0.250 &  5.81e-05 &  8.29e-05  &       1.18e-04 &  -9.396e+00  &       3.47e-01  &  3.39e+00  \\
0.300 &  6.96e-04 &  1.01e-03  &       1.60e-03 &  -6.856e+00  &       4.03e-01  &  4.21e+01  \\
0.350 &  4.96e-03 &  7.25e-03  &       1.19e-02 &  -4.877e+00  &       4.27e-01  &  5.94e+01  \\
0.400 &  2.34e-02 &  3.41e-02  &       5.58e-02 &  -3.326e+00  &       4.26e-01  &  5.65e+01  \\
0.450 &  8.16e-02 &  1.18e-01  &       1.88e-01 &  -2.090e+00  &       4.15e-01  &  5.02e+01  \\
0.500 &  2.27e-01 &  3.23e-01  &       5.07e-01 &  -1.085e+00  &       4.00e-01  &  4.43e+01  \\
0.600 &  1.11e+00 &  1.53e+00  &       2.28e+00 &  4.617e-01   &       3.62e-01  &  3.61e+01  \\
0.700 &  3.71e+00 &  4.93e+00  &       6.94e+00 &  1.623e+00   &       3.19e-01  &  3.13e+01  \\
0.800 &  9.76e+00 &  1.25e+01  &       1.68e+01 &  2.550e+00   &       2.76e-01  &  2.70e+01  \\
0.900 &  2.19e+01 &  2.72e+01  &       3.50e+01 &  3.320e+00   &       2.38e-01  &  2.19e+01  \\
\itshape 1.000 & \itshape 4.04e+01 & \itshape 5.03e+01 &  \itshape 6.25e+01 & \itshape 3.917e+00 & \itshape 2.17e-01 &  \textemdash \\
\itshape 1.250 & \itshape 1.06e+02 & \itshape 1.32e+02 &  \itshape 1.64e+02 & \itshape 4.881e+00 & \itshape 2.17e-01 &  \textemdash \\
\itshape 1.500 & \itshape 2.78e+02 & \itshape 3.46e+02 &  \itshape 4.30e+02 & \itshape 5.845e+00 & \itshape 2.17e-01 &  \textemdash \\
\itshape 1.750 & \itshape 4.71e+02 & \itshape 5.85e+02 &  \itshape 7.28e+02 & \itshape 6.372e+00 & \itshape 2.17e-01 &  \textemdash \\
\itshape 2.000 & \itshape 7.97e+02 & \itshape 9.91e+02 &  \itshape 1.23e+03 & \itshape 6.898e+00 & \itshape 2.17e-01 &  \textemdash \\
\itshape 2.500 & \itshape 1.58e+03 & \itshape 1.96e+03 &  \itshape 2.44e+03 & \itshape 7.582e+00 & \itshape 2.17e-01 &  \textemdash \\
\itshape 3.000 & \itshape 2.58e+03 & \itshape 3.21e+03 &  \itshape 3.99e+03 & \itshape 8.074e+00 & \itshape 2.17e-01 &  \textemdash \\
\itshape 3.500 & \itshape 3.76e+03 & \itshape 4.68e+03 &  \itshape 5.81e+03 & \itshape 8.451e+00 & \itshape 2.17e-01 &  \textemdash \\
\itshape 4.000 & \itshape 5.10e+03 & \itshape 6.34e+03 &  \itshape 7.88e+03 & \itshape 8.754e+00 & \itshape 2.17e-01 &  \textemdash \\
\itshape 5.000 & \itshape 8.12e+03 & \itshape 1.01e+04 &  \itshape 1.25e+04 & \itshape 9.220e+00 & \itshape 2.17e-01 &  \textemdash \\
\itshape 6.000 & \itshape 1.14e+04 & \itshape 1.42e+04 &  \itshape 1.76e+04 & \itshape 9.561e+00 & \itshape 2.17e-01 &  \textemdash \\
\itshape 7.000 & \itshape 6.89e-01 & \itshape 8.56e-01 &  \itshape 1.06e+00 & \itshape -1.550e-01 & \itshape 2.17e-01 & \textemdash  \\
\itshape 8.000 & \itshape 7.36e-01 & \itshape 9.15e-01 &  \itshape 1.14e+00 & \itshape -8.859e-02 & \itshape 2.17e-01 & \textemdash  \\
\itshape 9.000 & \itshape 7.78e-01 & \itshape 9.67e-01 &  \itshape 1.20e+00 & \itshape -3.349e-02 & \itshape 2.17e-01 & \textemdash  \\
\itshape 10.000& \itshape 8.15e-01 & \itshape 1.01e+00 & \itshape 1.26e+00  & \itshape 1.334e-02  & \itshape 2.17e-01 & \textemdash  \\
      \bottomrule   \bottomrule
    \end{tabular}
  }
\end{table*}

\noindent 0.04 eV. We assign a conservative factor of two uncertainty to this value to yield $\Gamma_{\gamma}$ $=$ 0.04(4) eV. Thus, the approximation made in Ref.~\cite{Gillespie:2017} is only applicable for low energy resonances below $E_{r}^{c.m.}$ $=$ 300 keV (see Table~\ref{tab:3}). To avoid relying on that assumption, Eqn.~\ref{eq:rates-ResStrength} is used to calculate the resonance strength when direct measurements are absent.\par
The $^{34}$S($p, \gamma$)$^{35}$Cl resonant reaction rate was calculated using the information provided in Table~\ref{tab:3} together with the Monte Carlo methods of Ref.~\cite{Longland:2010}. Where states have only been observed in the present study, upper limit proton partial widths have been assumed with $C^{2}S < 1$. For a few states where more than one assignment is possible for the present proton angular distributions, Table~\ref{tab:1} only shows our best assignment, whereas in Table~\ref{tab:3} we have considered all the possibilities from our measurement together with that of Ref.~\cite{Gillespie:2017}. There are two states at 6800-keV and 7213.8-keV,

\begin{figure}[ht]
 \begin{center}
   \includegraphics[width=0.5\textwidth]{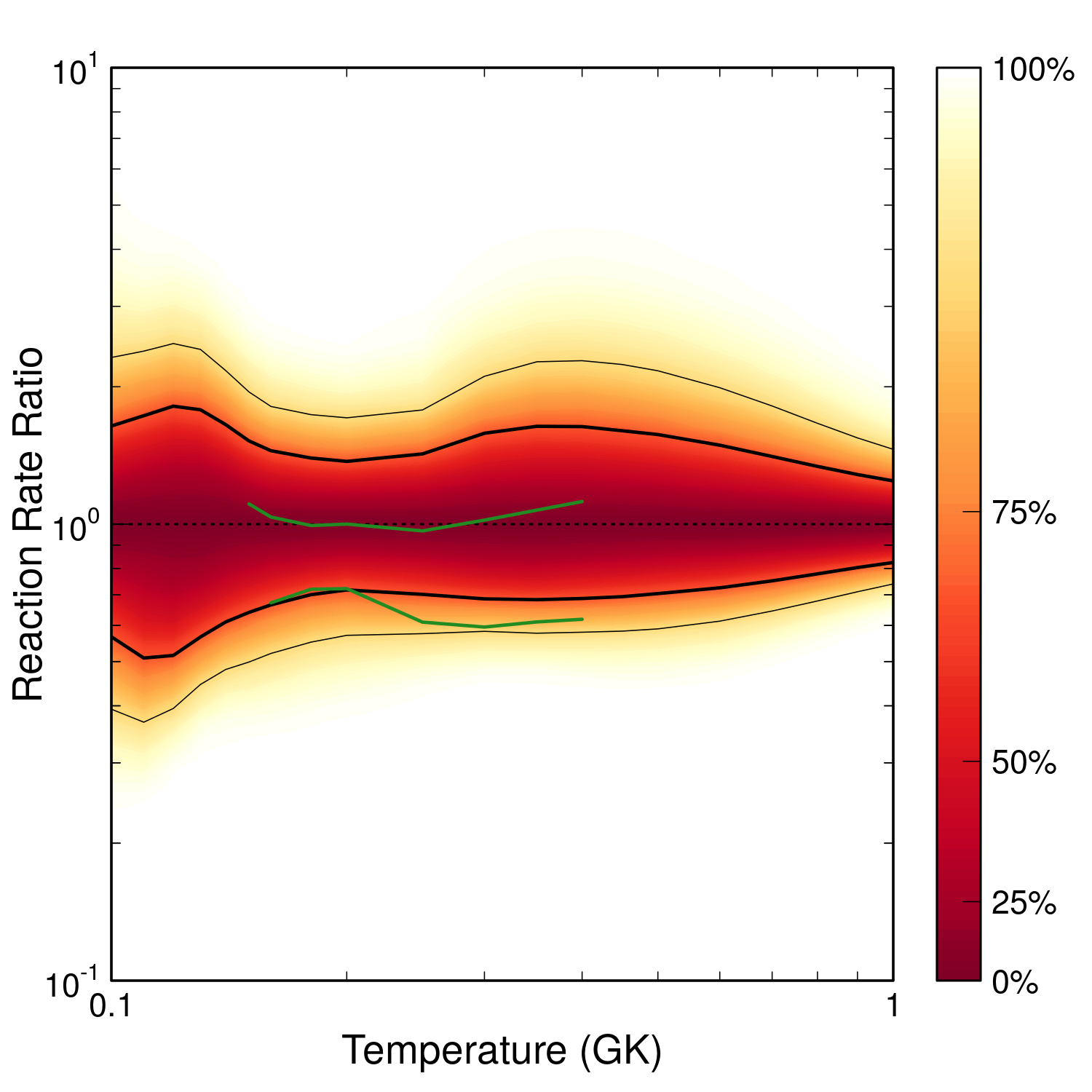}
 \end{center}
\caption{\label{GraphContour}(Color online) Rate uncertainties for the $^{34}$S($p, \gamma$)$^{35}$Cl reaction calculated using the resonance parameters presented in Tab.~\ref{tab:3}. The rates are normalized to the recommended rate so the recommended rate is at unity. The thick and thin solid lines correspond to the 1$\sigma$ and 2$\sigma$ uncertainty bands, respectively, with the color scale highlighting the continuous reaction rate probability distribution. The green (grey in print version) lines correspond to the ``high'' and ``low'' reaction rates presented by Ref.~\cite{Gillespie:2017}.}
\end{figure}

\noindent where no spin/parity information is available from the literature. The maximum contribution of these states to the reaction rate is calculated here with the presumption that these are s-wave resonances ($J^{\pi}$ $=$ 1/2$^{+}$). Since the 6800-keV state is inside the Gamow window for the nova temperature regime, we have also considered its next highest rate contribution if it is a p- or a d-wave resonance ($J^{\pi}$ $=$ 3/2$^{-}$ and 3/2$^{+}$, respectively). Six states at $E_{r}^{c.m.} = 56.7$ keV, 121.1 keV, 175.2 keV, 390.2 keV, 495.7 keV, and 807.2 keV have ambiguous spin parity assignments, so we use the method outlined in Ref.~\cite{Mohr:2014} to sample the possibilities with a 50\% probability for each $\ell$-value (angular momentum transfer). Finally, in the Monte Carlo sampling, a Porter-Thomas distribution is assumed with a mean single-particle reduced width of $\theta_{sp}^{2} = 4.5 \times 10^{-3}$ according to the findings of Ref.~\cite{Pogrebnyak:2013}. The final reaction rates are presented in Table~\ref{tab:pgRate}. Those shown in italics denote Hauser-Feshbach reaction rates from the code TALYS~\cite{Goriely:2008} that have been normalized to the experimental rate at 1.0 GK. This matching temperature was found using the methods outlined in Ref.~\cite{Newton:2008}.\par
The uncertainty band for the reaction rate is shown in Fig.~\ref{GraphContour}. Here, the reaction rate uncertainty bands have been normalized to the median, recommended rate at unity. Also shown is the so-called ``high'' and ``low'' rates from Ref.~\cite{Gillespie:2017}. Over the temperature range of 0.1 -- 0.4 GK, the reaction rates presented here are in agreement with those from Ref.~\cite{Gillespie:2017}. However, the reaction rate uncertainty band presented here is larger than that presented in Ref.~\cite{Gillespie:2017} owing to our treatment of the uncertainties in all inputs to the reaction rate calculation, including the resonance energy uncertainty, which enters into the penetrability calculation for proton partial widths. We have also included $\gamma$-ray partial widths, which affect the rate calculation at higher temperatures where the assumption made in Ref.~\cite{Gillespie:2017} is no longer valid. The present high to low reaction rate ratio, which is a measure of the rate uncertainty, peaks at a factor of 3.5 at 0.12 GK. In comparison, those rates from Ref.~\cite{Gillespie:2017} differ by less than a factor of 2.\par

\begin{figure}[ht]
 \begin{center}
   \includegraphics[width=0.5\textwidth]{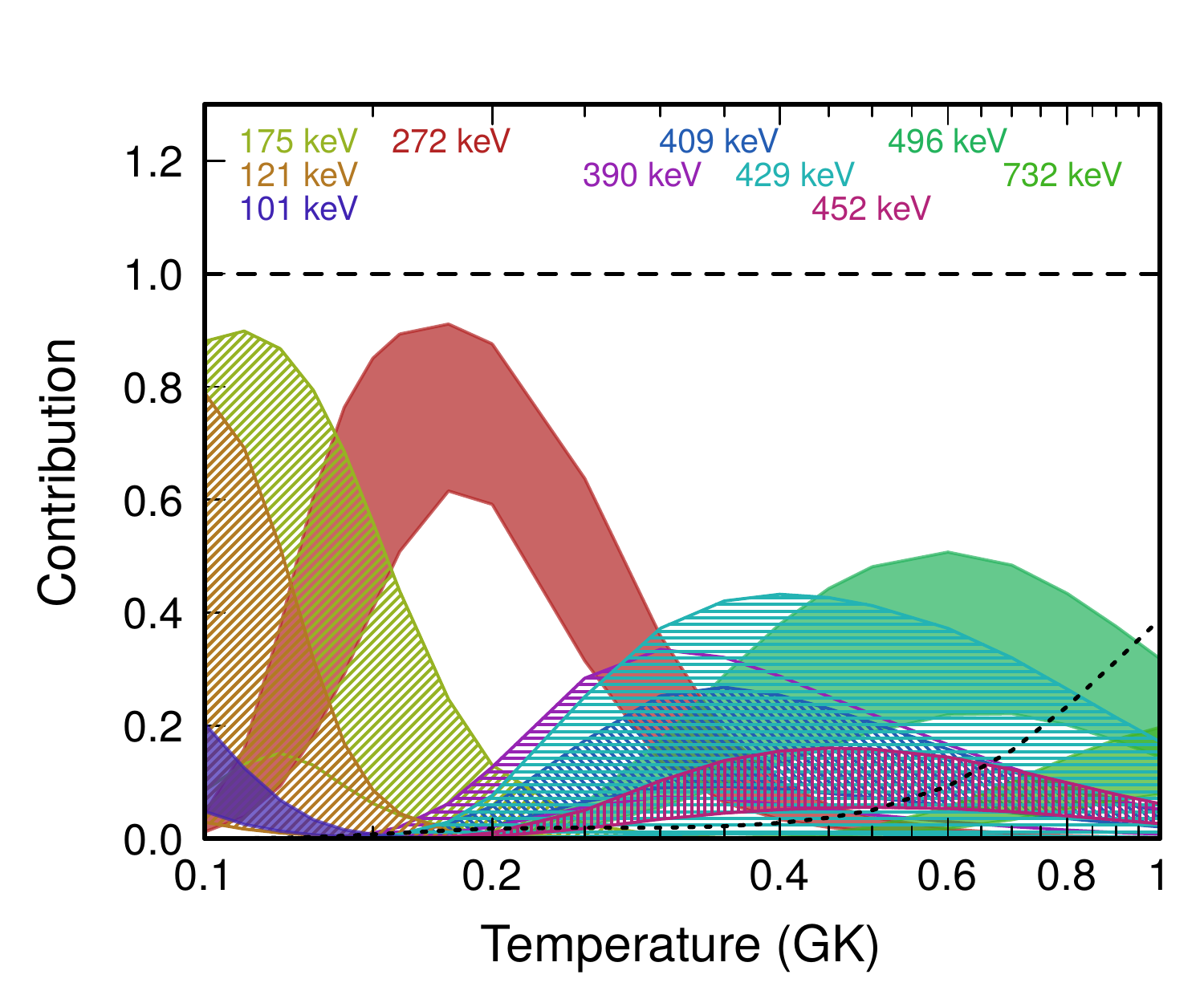}
 \end{center}
\caption{\label{GraphContribution}(Color online) Resonance contributions to the total reaction rate. Each color band signifies a single narrow resonance and its contribution to the reaction rate. A finite thickness to these lines denotes those resonances which may contribute significantly to the rate, or may only contribute in a minor way. For example, all three resonances at $E_{r}^{c.m.} = $ 390 keV, 409 keV, and 429 keV are not known well enough to determine which dominates the reaction rate at 250 MK. The dotted line represents the aggregate contribution of higher lying resonances not significant for the nova temperature regime.}
\end{figure}

Figure~\ref{GraphContribution} shows the contributing resonances over the temperature range of interest. The $^{35}$Cl excited states that significantly contribute to the present $^{34}$S($p, \gamma$)$^{35}$Cl reaction rate over the nova temperature regime are at $E_{r}^{c.m.}$ $=$ 121.1 keV, 175.2 keV, 272.2 keV, 390.2 keV, 409.2 keV and 429.2 keV with the latter 3 resonances becoming more important at $T$ $\gtrapprox$ 0.25 GK. Although due to their large proton width uncertainties, it is not possible to unambiguously identify which ones matter the most.\par
It is worth mentioning that the 272.2-keV resonance corresponding to the 6643-keV state, observed for the first time in Ref.~\cite{Gillespie:2017}, has a significant effect on the present reaction rate. Upon inspection of Fig.~2 in Ref.~\cite{Gillespie:2017}, this state is in the vicinity of a background peak in that region. However, the presented angular distribution of the outgoing deuterons corresponding to this state (see Fig.~3 in Ref.~\cite{Gillespie:2017}) reveals that it was observed at 5 angles. Gillespie {\em et al}.~\cite{Gillespie:2017} have considered an unambiguous $\ell$ $=$ 1 ($^{3}$He, $d$) transfer for the 6643-keV state. However, the theoretical DWBA curve in their Fig.~3 is not properly scaled to the data for an $\ell$ $=$ 1 transfer. This state remained unobserved in our measurement and those of Refs.~\cite{Bisoi:2013,Chipps:2017}. Moreover, it is near an energy window where a doublet is expected~\cite{Goss:1973}, and we have observed both those states at 6662.2(19) keV and 6677(3) keV. If we remove the 6643-keV state from our rate calculation to examine the significance of its effect on the present rate, our rate becomes smaller than that of Ref.~\cite{Gillespie:2017} by up to a factor of 3.2 over the nova temperature regime. Without doubt, further studies should be performed to confirm the existence of this state by an independent measurement and to examine the $^{35}$Cl states in the excitation energy window of 6.6 to 6.7 MeV.\par
The 121.1-keV, 175.2-keV and 409.2-keV resonances were measured in both the present work and Ref.~\cite{Gillespie:2017}. The 390.2-keV resonance has been measured by Ref.~\cite{Gillespie:2017} (although it still has an ambiguous spin parity assignment). Finally, the resonance at 429.2 keV has also been observed in the present work but is treated as an upper limit because we do not have enough proton angular distribution data to perform a reliable DWBA calculation. These resonances should be the focus of further study to determine their properties unambiguously.

\section{\label{Conclusions}Conclusions}

This study presented a charged-particle spectroscopy experiment using the Enge split-pole spectrograph at TUNL to study the excitation energy range of 6 -- 7 MeV in $^{35}$Cl via the $^{32}$S($\alpha, p$)$^{35}$Cl reaction at $E_{\alpha}$ $=$ 21 MeV. Properties of the $^{35}$Cl proton resonances in this energy window determine the $^{34}$S($p, \gamma$)$^{35}$Cl reaction rate over the temperatures characteristic of explosive hydrogen burning in novae. A precise knowledge of this rate, in turn, may help discriminate between presolar grains of nova (oxygen-neon) origin and those of other stellar sources, such as type II supernovae.\par
The $^{35}$Cl excitation energies measured here mostly agree within 1 -- 2$\sigma$ with the results of previous experiments~\cite{Chen:2011,Gillespie:2017}. There are only two states observed in this study at $E_{x}$ $=$ 5531(4) keV and 5731(3) keV (the 2-keV systematic uncertainty is also considered), which are in disagreement with the previously measured values~\cite{Chen:2011} beyond 2$\sigma$. However, both these states are proton bound and do not contribute to the $^{34}$S($p, \gamma$)$^{35}$Cl reaction rate at the nova temperature regime. In addition, another state is observed at 6662(3) keV, considering the 2-keV systematic uncertainty, whose energy agrees within 1$\sigma$ with the result of the measurement of Ref.~\cite{Bisoi:2013} and the latest evaluated value~\cite{Chen:2011} at 6656(3) but is in disagreement with $E_{x}$ $=$ 6643(2) keV measured in Ref.~\cite{Gillespie:2017}. Our derived spin and parity for the 6662-keV state does not match that of the 6643-keV state either. We have therefore, considered these as nonidentical states. The 6643-keV state dominates the $^{34}$S($p, \gamma$)$^{35}$Cl reaction rate from $\sim$ 0.14 GK to $\sim$ 0.25 GK, and if we remove it from our rate calculation to probe its effect, the present rate decreases by a factor of 3.2 at these temperatures. We thereby recommend future measurements to investigate this excitation energy region, particularly because this is a region where a doublet is expected~\cite{Goss:1973}. We have firmly assigned the spin parity of the higher energy state in the doublet at 6677(3) keV as 1/2$^{+}$, confirming the assumption made for the first time in Ref.~\cite{Chipps:2017} concerning the $J^{\pi}$ value of this state. In the present work, the strengths of the 288.3- and 306.2-keV resonances, corresponding to the 6659.1- and 6677-keV states, respectively, are treated as upper limits. Their average contributions to the $^{34}$S($p, \gamma$) reaction rate are too small to be shown on Figure~\ref{GraphContribution}. If instead we adopt the resonance strength for the 6677-keV state from Ref.~\cite{Gillespie:2017} without treating it as an upper limit, then this state has an effect on the rate up to about 20\% at 0.2 GK.\par
Ten new states were discovered in the measurement of Ref.~\cite{Gillespie:2017}. With the exception of the 6329-, 6643-, and 6823-keV states, not observed here, we have confirmed the existence of all the other ones.\par
The theoretical angular distributions of the $^{32}$S($\alpha, p$) reaction were computed via DWBA calculations. The potential contribution of the compound nucleus to the ($\alpha, p$) reaction is beyond the scope of this work and was not considered here. To improve upon the optical potential model used for the DWBA calculations, a $^{32}$S $+$ $\alpha$ elastic scattering measurement was also performed at $E_{\alpha}$ $=$ 21 MeV. The present spins and parities derived for the $^{35}$Cl states of interest to nova nucleosynthesis mostly agree with the values found in the literature~\cite{Chen:2011,Gillespie:2017}. However, there are some cases, e.g., the 6428.6-, 6662.2-, and 6863.1-keV states, where the present $J^{\pi}$ assignments are in disagreement with those of Ref.~\cite{Gillespie:2017}. We have also tentatively assigned $J^{\pi}$ values to five excited states of $^{35}$Cl for the first time.\par
The $^{34}$S($p, \gamma$)$^{35}$Cl reaction rate at the nova temperatures was recalculated based on the Monte Carlo techniques developed in Refs.~\cite{Longland:2010,Iliadis:2010a,Iliadis:2010b}. Over the temperature range of interest, the present rate is consistent with that of Ref.~\cite{Gillespie:2017}. With respect to the latter, the uncertainty in the present reaction rate is larger due to properly considering the uncertainties in all inputs to the reaction rate calculation. The ambiguities in the properties of six resonances at $E_{r}^{c.m.}$ $=$ 121.1 keV, 175.2 keV, 272.2 keV, 390.2 keV, 409.2 keV and 429.2 keV does not allow us to exclusively isolate the one that plays the most significant role in determination of the $^{34}$S($p, \gamma$)$^{35}$Cl reaction rate at 0.1 $\leq$ $T$ $\leq$ 0.4 GK. Thus further study is warranted.\par
Once these discrepancies are resolved, one can obtain a more reliable theoretical $^{34}$S/$^{32}$S ratio that can be compared with that obtained experimentally from presolar grains to more reliably identify if they originated from an oxygen-neon nova.

\section*{\label{Acknowledgment}ACKNOWLEDGEMENTS}

The authors thank the TUNL technical staff for their contributions. This material is based upon work supported by the U.~S.~Department of Energy, Office of Science, Office of Nuclear Physics, under Award Number DE-SC0017799 and under Contract No.~DE-FG02-97ER41041.


\bibliographystyle{apsrev}
\bibliography{References}
\end{document}